\documentclass[a4paper,11pt]{article}
\usepackage{amsmath, amssymb, graphics, graphicx}
\usepackage{color}
\usepackage{longtable}
\usepackage{subfigure}
\usepackage{multirow}

\newcommand{\mathsym}[1]{{}}

\newcommand{\F}{\mathcal{F}}
\newcommand{\I}{\mathrm{Im}}
\newcommand{\R}{\mathrm{Re}}
\newcommand{\N}{\mathcal{N}}
\newcommand{\rmd}{\mathrm{d}}
\newcommand{\rmi}{\mathrm{i}}
\newcommand{\rme}{\mathrm{e}}
\newcommand{\rmg}{\mathrm{g}}
\newcommand{\rmh}{\mathrm{h}}
\newcommand{\rmp}{\mathrm{p}}
\newcommand{\rmq}{\mathrm{q}}
\newcommand{\rmX}{\mathrm{X}}
\newcommand{\rmY}{\mathrm{Y}}
\newcommand{\rmZ}{\mathrm{Z}}
\newcommand{\SO}{\mathop{\rm SO}}
\newcommand{\SL}{\mathop{\rm SL}}
\newcommand{\Sp}{\mathop{\rm Sp}}
\newcommand{\SU}{\mathop{\rm SU}}
\newcommand{\U}{\mathop{\rm U}}

\renewcommand{\theequation}{\thesection.\arabic{equation}}
\csname @addtoreset\endcsname{equation}{section}

\begin{document}
\begin{titlepage}
\begin{flushright}
DFTT 13/2008
\end{flushright}
\vskip 1.5cm
\begin{center}
{\LARGE \bf  On full-fledged supergravity cosmologies \\ \vskip 0.2cm and their Weyl group asymptotics} \\
 \vfill
{\large Pietro Fr{\'e}, Jan Rosseel} \\
\vfill {
Dipartimento di Fisica Teorica, Universit{\`a} di Torino, \\
$\&$ INFN -
Sezione di Torino\\
via P. Giuria 1, I-10125 Torino, Italy}
\end{center}
\vfill
\begin{abstract}
We consider a class of cosmological solutions of $d=4$, $\mathcal{N}=2$ supergravity theories coupled to vector multiplets. The solutions result from performing a compactification to three dimensions, where the theory reduces to a symmetric space sigma model coupled to gravity and where the resulting equations of motion are integrable. We describe in detail how the three-dimensional solutions can be uplifted to four dimensions again. The four-dimensional cosmologies are generically characterized by an algebra of translational isometries that is of Heisenberg type. We give explicit examples of these cosmologies for the $S$-$T$-$U$ model and comment also on their uplift to 10 dimensions by interpreting them as solutions of a truncation of type IIB supergravity on a $K3 \times T^2/\mathbb{Z}_2$ orientifold.
\end{abstract}
\vspace{2mm} \vfill \hrule width 3.cm
{\footnotesize \noindent
e-mails: \{fre, rosseel\}@to.infn.it}
\end{titlepage}
\addtocounter{page}{1}
 \tableofcontents{}
\newpage
\section{Introduction} \label{sec:intro}

In the last years there has been a lot of interest in cosmological applications of string theory and supergravity. In this context, attention has been devoted to cosmological solutions of supergravity theories, in particular with regard to questions pertaining to inflation, brane-world scenario's, accelerating universes and the like (see e.g. \cite{HenryTye:2006uv}, \cite{Cline:2006hu}, \cite{Kallosh:2007ig}, \cite{Linde:2007fr}, \cite{Burgess:2007pz}, \cite{McAllister:2007bg} for some reviews). The main feature of supergravity cosmology is the presence of extra dimensions and corresponding scalar fields. In relation with them, there arises the interesting phenomenon which goes under the name of the cosmic billiard paradigm. This refers to the generic feature that cosmological scale factors undergo repeated changes, resulting in the fact that certain directions start decreasing after a regime in which they increased and vice versa (see e.g. \cite{Damour:2002et}, \cite{Henneaux:2007ej} and references therein). In a series of papers \cite{Fre:2003ep}, \cite{Fre:2003tg}, \cite{Fre':2005sr}, \cite{Fre:2005bs}, \cite{Fre':2007hd}, it was shown that exact supergravity solutions displaying this billiard phenomenon, can be systematically derived relying on the algebraic structure of the duality algebras underlying supergravity. It has even been proven that the supergravity field equations, reduced to only time-dependence, constitute an integrable system and the general integral depending on all integration constants can in fact be constructed in an algorithmic manner. 

The main point of this integration relies on the strategy of reducing supergravity to low dimensions $d \leq 3$, where all bosonic degrees of freedom are represented by scalar fields, effectively reducing the theory to a sigma model. In case this sigma model is a symmetric space, the resulting equations of motion are integrable. This reduction does not lose generality in finding solutions as long as we are interested only in time-dependence, as is the goal of cosmology. So far, the discussion of the inversion of this procedure, namely the reconstruction (or oxidation) of higher-dimensional cosmological supergravity solutions obtained from the above method has been only briefly touched upon. A thorough investigation of the properties of these cosmologies and the visualization of what the billiard phenomenon actually means in higher, physical dimensions has not been presented. This is the main goal of the present paper.

In this paper, we will focus mainly on oxidation to four dimensions; we can however apply similar techniques to oxidize to higher dimensions, if possible. Indeed, we will also present an example in which we oxidize to ten dimensions. Anticipating one of the results of this paper, we can already mention that the higher-dimensional ($d\geq 4$) cosmologies, that are accessible to our general integration method, correspond to cosmologies in which the algebra of translational isometries is non-trivial. At this point, it is useful to note the intrinsic meaning of considering time-dependent solutions. Cosmologies are usually characterized by asserting the homogeneity of space, leading to the existence of a certain number of translational Killing vectors. The algebra satisfied by these translations can be abelian ("flat universes") or non-abelian. The higher-dimensional cosmologies we find are characterized by an algebra of translation generators $\{T_i$, $c\}$ that is of Heisenberg type:
\begin{equation} \label{heisalg}
[T_i, T_j] = c \,, \qquad [c, T_i] = 0 \,.
\end{equation}
Such algebras represent a mild deformation of flat universes. To make this more explicit, we mention one of the examples considered in this paper, that corresponds to type IIB supergravity compactified on a $K3 \times T^2/\mathbb{Z}_2$ orientifold. We will show that using our method, we can find an exact solution of the supergravity field equations in which the metric assumes the form:
\begin{eqnarray} \label{form10dmetricintro}
\rmd s^2_{10d} & = & A(t)^{-1/2} \left[-\frac{B^4(t)}{\Delta(t)} \rmd t^2 + \frac{B^2(t)}{\Delta(t)}\left(\rmd x^2 + \rmd y^2\right) + \Delta(t) \left(\rmd z + \alpha \Omega\right)^2 \right] \nonumber \\ & & + A(t)^{-1/2} \left[C^2(t) \rmd u^2 + D^2(t) \rmd v^2 + E^2(t) \rmd u \rmd v \right] + A(t)^{1/2} \rmd s^2_{K3} \,.
\end{eqnarray}
In the above metric, $t$, $x$, $y$ and $z$ are coordinates of the four-dimensional space-time, $u$ and $v$ denote the $T^2$-coordinates and $\rmd s^2_{K3}$ represents the $K3$-metric. The metric depends on a constant $\alpha$ and the time-dependent functions $A(t), \cdots, E(t), \Delta(t)$, that will be determined later on. The one-form $\Omega$ is given by
\begin{equation}
\Omega = - \frac{y}{2} \rmd x + \frac{x}{2} \rmd y \,.
\end{equation}
The above metric (\ref{form10dmetricintro}) exhibits a certain number of translational isometries that act as ordinary translations on the $T^2$-coordinates and act as follows on $x$, $y$ and $z$:
\begin{eqnarray}
\delta x & = & a\,, \nonumber \\
\delta y & = & b\,, \nonumber \\
\delta z & = & c - \frac{\alpha}{2} a y + \frac{\alpha}{2} b x \,.
\end{eqnarray}
These translations indeed close an algebra of Heisenberg type (\ref{heisalg}).

In this paper, we focus in particular on oxidation to $\mathcal{N} = 2$ ungauged theories in four dimensions with vector multiplets, which often can be traced back to compactifications of superstring theory on Calabi-Yau manifolds. For this class of theories, the bosonic Lagrangian depends on the metric and a certain number of scalars and vector fields. The cosmological solutions of such a Lagrangian which we are able to construct by our method, are of the following form:
\begin{equation} \label{explsol4dmetric}
\rmd s^2_{4d} = -\frac{B^4(t)}{\Delta(t)} \rmd t^2 + \frac{B^2(t)}{\Delta(t)}\left(\rmd x^2 + \rmd y^2\right) + \Delta(t) \left(\rmd z + \alpha \Omega\right)^2\,.
\end{equation}
The integration algorithm not only allows to determine the explicit analytic expression for the involved time-dependent functions $\Delta(t)$, $B(t)$; it also determines the explicit value of the constant $\alpha$ and the explicit time-dependent expressions for the scalars and the vectors. The solutions depend on a complete set of integration constants , given in terms of the number $n_V$ of vector multiplets as:
\begin{equation}
\mathrm{number \ of \ integration \ constants} = 8 n_V + 8 \,.
\end{equation}
Our goal was furthermore to enlighten the billiard features exhibited by solutions of this type. From the general results of previous papers we know that asymptotically at $t = \pm \infty$ the solutions tend to simplify and have the general appearance of Kasner metrics. The asymptotic metrics at $t=-\infty$ and $t=+\infty$ furthermore turn out to be related to each other via the action of the Weyl group of the isometry group of the sigma model that appears after reduction to three dimensions. We will therefore explicitly show how the Weyl group acts on solutions of the type (\ref{explsol4dmetric}) above. 

The organization of this paper is as follows. In section \ref{sec:cosmsol}, we will give a theoretical discussion on how we can combine previous results on obtaining time-dependent solutions of lower-dimensional supergravities, with the process of dimensional reduction/oxidation, to obtain cosmologies of $\mathcal{N}=2$, $d=4$ supergravity. We will describe the dimensional reduction from four to three dimensions in detail and summarize the integration algorithm that can be used to obtain time-dependent solutions of the three-dimensional theory. After that, we will give a detailed account on how one can reinterpret the latter solutions as complete solutions of the original four-dimensional supergravity. After this theoretical discussion, we will put the theory into practice in the context of a specific $\mathcal{N}=2$, $d=4$ supergravity in section \ref{sec:examples}. After a general discussion on the choice of supergravity, we will give three examples of exact cosmological solutions of this model. For each of these examples, we will give explicit solutions, indicate some properties and highlight their asymptotic behavior and its relation with the relevant Weyl group. The supergravity model of section \ref{sec:examples} can be obtained by performing a $K3\times T^2/\mathbb{Z}_2$ orientifold compactification of type IIB supergravity. This allows us to reinterpret the solutions of section \ref{sec:examples} as solutions of ten-dimensional supergravity. This step will be taken in section \ref{sec:10d}. Finally, we end the paper with some appendices containing technical results and conventions.

\section{Cosmological solutions of $d=4$, $\mathcal{N}=2$ supergravity} \label{sec:cosmsol}

This section describes the method used to construct the class of four-dimensional cosmological solutions, discussed in this paper. We will exhibit the method as a three step process. The first step consists in studying the dimensional reduction of $d=4$, $\mathcal{N}=2$ supergravity to a non-linear sigma model coupled to gravity in $d=3$. In order to establish our notation, we will describe this reduction in detail. This step will be described in section \ref{ssec:red}. In the second step, we obtain time-dependent solutions of the reduced three-dimensional theories. In order to find these solutions, we use the techniques developed in \cite{Fre:2003ep,Fre':2005sr,Fre:2005bs,Fre':2007hd}. For the benefit of the reader, we have shortly summarized these techniques in section \ref{ssec:obtsol}. Finally, in the third step, the three-dimensional time-dependent solutions are reinterpreted as cosmological solutions of four-dimensional $\mathcal{N}=2$ supergravity. In order to do this, one needs a precise mapping between four-dimensional and three-dimensional fields. This mapping can be constructed by relying on the discussion of section \ref{ssec:red} and on group theoretical arguments. This will be explained in more detail in section \ref{ssec:uplift}.

\subsection{Step 1 : Dimensional reduction of $d=4$, $\mathcal{N}=2$ supergravity} \label{ssec:red}

In this section, we describe the dimensional reduction of four-dimensional supergravity with 8 supercharges to three dimensions in detail. Although this dimensional reduction is well-known (see e.g. \cite{Cecotti:1988qn,Ferrara:1989ik}), we prefer to repeat it here both in order to establish notation as well as to highlight some steps that will be useful later on.

The class of cosmological solutions we will consider in this paper, corresponds to solutions of $d=4$, $\mathcal{N}=2$ supergravity coupled to $n_V$ vector multiplets. The bosonic Lagrangian for the theories under consideration is then given by \cite{deWit:1985px, Andrianopoli:1996vr,Andrianopoli:1997cm}:
\begin{eqnarray} \label{4dlagr}
\hat{e}^{-1} \hat{\mathcal{L}}_{4d} & = & \frac{1}{2} \hat{R} - g_{\alpha \bar{\beta}} \partial_{\hat{\mu}} w^\alpha \partial^{\hat{\mu}} \bar{w}^{\bar{\beta}} + \frac{1}{4} \left( \I \N_{IJ}\right) \hat{\F}^I_{\hat{\mu} \hat{\nu}} \hat{\F}^{J \hat{\mu} \hat{\nu}} \nonumber \\
& & -\frac{1}{8} \left(\R \N_{IJ} \right) \hat{e}^{-1} \varepsilon^{\hat{\mu} \hat{\nu} \hat{\rho} \hat{\sigma}} \hat{\F}^I_{\hat{\mu} \hat{\nu}} \hat{\F}^J_{\hat{\rho} \hat{\sigma}} \,.
\end{eqnarray}
In the above Lagrangian, we have denoted by $w^\alpha$, $\alpha = 1, \ldots, n_V$ the $n_V$ complex scalars in the vector multiplets. They span a non-linear sigma model where the target space is a special K\"ahler manifold with metric $g_{\alpha \bar{\beta}}$. The Lagrangian (\ref{4dlagr}) also contains $n_V + 1$ vector fields $\hat{A}^I_{\hat{\mu}}$, $I = 0, \ldots, n_V$, whose field strengths are denoted by $\hat{\mathcal{F}}^I_{\hat{\mu} \hat{\nu}} = 2\, \partial_{[\hat{\mu}} \hat{A}^I_{\hat{\nu}]}$. The so-called period matrix $\mathcal{N}_{IJ}$, that determines the terms involving the vector fields is a symmetric, complex matrix that depends on the scalar fields $w^\alpha$ (as well as their complex conjugates $\bar{w}^{\bar{\alpha}}$) and whose imaginary part is negative definite. Note that we have adopted a notation in which four-dimensional indices are denoted with a hat. For the fields themselves a similar notation is used, except for the scalar fields, as their reduction is trivial.

Denoting the compact coordinate by $z$ and using ordinary, non-hatted indices for the non-compact three-dimensional coordinates, the reduction then proceeds by taking the usual Kaluza-Klein ansatz for the line element:
\begin{equation} \label{redmetric}
\rmd s^2_{4d} = \Delta^{-1} \rmd s^2_{3d} + \Delta \left( \rmd z + A^{(0)}_\mu \rmd x^\mu\right)^2 \,.
\end{equation}
For the vector fields on the other hand, the following ansatz is used:
\begin{eqnarray} \label{redvectors}
\hat{A}^I_\mu & = & A^I_\mu + \chi^I A^{(0)}_\mu \,, \nonumber \\
\hat{A}^I_z  & = & \chi^I \,.
\end{eqnarray}
Using the above formulas, one obtains the following Lagrangian in three dimensions:
\begin{eqnarray} \label{3dlagrundual}
e^{-1} \mathcal{L}_{3d} & = & \frac{1}{2} R - \frac{1}{4 \Delta^2} \partial_\mu \Delta \partial^\mu \Delta - \frac{1}{8} \Delta^2 \F^{(0)}_{\mu \nu} \F^{(0) \mu \nu} \nonumber \\ & & - g_{\alpha \bar{\beta}} \partial_\mu w^\alpha \partial^\mu \bar{w}^{\bar{\beta}} + \frac{1}{2} \left(\I \N_{IJ}\right) \Delta^{-1} \partial_\mu \chi^I \partial^\mu \chi^J \nonumber \\ & & + \frac{1}{4} \left( \I \N_{IJ} \right) \Delta \left( \F^I_{\mu \nu} + \chi^I \F^{(0)}_{\mu \nu}\right) \left( \F^{J \mu \nu} + \chi^J \F^{(0)}_{\mu \nu} \right) \nonumber \\ & & -\frac{1}{2} \left(\R \N_{IJ}\right) \varepsilon^{\mu \nu \rho} e^{-1} \left( \F^I_{\mu \nu} + \chi^I \F^{(0)}_{\mu \nu} \right) \partial_\rho \chi^J \,.
\end{eqnarray}
The above Lagrangian still contains the three-dimensional vector fields. In three dimensions however, vectors are dual to scalar fields; thus one can obtain a Lagrangian where only scalar fields and the metric are present. In order to dualize the vectors, we add the Lagrange multipliers $\sigma_I$ and $\omega$ to $\mathcal{L}_{3d}$ by writing:
\begin{equation} \label{lagrmult}
\mathcal{L}_{\mathrm{mult}} = \frac{1}{2} \varepsilon^{\mu \nu \rho} \F^I_{\nu \rho} \partial_\mu \sigma_I - \frac{1}{4} \varepsilon^{\mu \nu \rho} \F^{(0)}_{\nu \rho} \partial_\mu\left(\omega - \chi^I \sigma_I \right) \,.
\end{equation}
Considering $\mathcal{L} = \mathcal{L}_{3d} + \mathcal{L}_{\mathrm{mult}}$ and imposing 
\begin{equation}
\frac{\delta \mathcal{L}}{\delta \F^I_{\mu \nu}} = 0 \,, \qquad \frac{\delta \mathcal{L}}{\delta \F^{(0)}_{\mu \nu}} = 0 \,,
\end{equation}
the following duality relations are obtained:
\begin{eqnarray} \label{dualityrels}
\F^{(0)}_{\mu \nu} & = & - \frac{1}{\Delta^2} e \varepsilon_{\mu \nu \rho} \left[ \partial^\rho \omega + \chi^I \stackrel{\leftrightarrow}{\partial^\rho} \sigma_I \right] \,, \nonumber \\
\F^{I}_{\mu \nu} & = & \frac{e}{\Delta^2} \varepsilon_{\mu \nu \rho} \bigg[ \Delta \left(\I \N \right)^{-1|IJ} \left( (\R \N)_{JK} \partial^\rho \chi^K - \partial^\rho \sigma_J \right) \nonumber \\Ê& & +\  \chi^I \left( \partial^\rho \omega + \chi^J \stackrel{\leftrightarrow}{\partial^\rho} \sigma_J \right) \bigg] \,.
\end{eqnarray}
Using these relations in $\mathcal{L}$, we obtain a three-dimensional Lagrangian that is solely expressed in terms of the metric and the scalar fields:
\begin{eqnarray} \label{3dlagrdual}
e^{-1}\tilde{\mathcal{L}}_{3d} & = & \frac{1}{2} R - \frac{1}{4 \Delta^2} \partial_\mu \Delta \partial^\mu \Delta - g_{\alpha \bar{\beta}} \partial_\mu w^\alpha \partial^\mu \bar{w}^{\bar{\beta}} \nonumber \\ & & + \frac{1}{2} \Delta^{-1} \left(\I \N\right)_{IJ} \partial_\mu \chi^I \partial^\mu \chi^J \nonumber \\ & & - \frac{1}{4 \Delta^2} \left( \partial_\mu \omega + \chi^I \stackrel{\leftrightarrow}{\partial_\mu} \sigma_I \right) \left( \partial^\mu \omega + \chi^J \stackrel{\leftrightarrow}{\partial^\mu} \sigma_J \right) \nonumber \\ & & + \frac{1}{2 \Delta} \left(\I \N\right)^{-1|IJ} \left[\left(\R \N\right)_{IM} \partial^\mu \chi^M - \partial^\mu \sigma_I \right]  \nonumber \\ & & \qquad \qquad \left[\left(\R \N\right)_{JN} \partial_\mu \chi^N - \partial_\mu \sigma_J \right] \,.
\end{eqnarray}
Summarizing, upon dimensional reduction, the four-dimensional fields lead to scalars in three dimensions according to the following scheme:
\begin{eqnarray} \label{reducscheme}
w^\alpha & \longrightarrow & w^\alpha \qquad \qquad 2n_V \nonumber \\
\hat{A}^I_{\hat{\mu}} & \longrightarrow & \left\{\begin{array}{l} \chi^I \ \ \, \qquad n_V+1 \\ \sigma_I \ \ \, \qquad n_V+1 \end{array} \right. \nonumber \\
\hat{g}_{\hat{\mu} \hat{\nu}} & \longrightarrow &  \left\{\begin{array}{l} \Delta \ \ \, \, \qquad 1 \\ \omega \ \ \ \, \qquad 1\end{array} \right.
\end{eqnarray}
In the last column of this scheme, we have mentioned the real number of degrees of freedom represented by each entity. The $\sigma_I$ and $\omega$ scalars are respectively obtained from dualization of the three-dimensional vectors $A^I_\mu$ and $A^{(0)}_\mu$. From the above scheme, one notes that in total $4(n_V + 1)$ scalars appear in the three-dimensional Lagrangian. The three-dimensional Lagrangian takes the form of a non-linear sigma model coupled to gravity. It is well-known that the target space manifold of this non-linear sigma model is quaternionic-K\"ahler \cite{Cecotti:1988qn,Ferrara:1989ik}. 

The quaternionic-K\"ahler manifolds, obtained by dimensional reduction of $d=4$, $\mathcal{N}=2$ supergravity coupled to vector multiplets are in fact called special quaternionic-K\"ahler manifolds. The class of special quaternionic-K\"ahler manifolds is rather general, however an interesting subclass of it is given by manifolds that are also homogeneous. They are thus given by coset spaces $\mathcal{G}/\mathcal{H}$, where $\mathcal{G}$ corresponds to the isometry group of the manifold and $\mathcal{H}$ to its isotropy group. A proper subclass of these homogeneous quaternionic-K\"ahler manifolds is given by those that are also symmetric spaces. In this case, the Lie algebra $\mathbb{G}$ of $\mathcal{G}$ admits a decomposition in terms of the Lie algebra $\mathbb{H}$ of $\mathcal{H}$ and the orthogonal complement $\mathbb{K}$:
\begin{equation} \label{GdecompHK}
\mathbb{G} = \mathbb{H} \oplus \mathbb{K} \,,
\end{equation}
such that the following commutation relations hold:
\begin{eqnarray} \label{defsymmspace}
\big[\mathbb{H},\mathbb{H}\big] & \subset & \mathbb{H} \,, \nonumber \\
\big[\mathbb{H},\mathbb{K}\big] & \subset & \mathbb{K} \,, \nonumber \\
\big[\mathbb{K},\mathbb{K}\big] & \subset & \mathbb{H} \,.
\end{eqnarray}
The Lie algebra $\mathbb{G}$ is an appropriate real form of a complex Lie algebra $\mathbb{G}_{\mathbb{C}}$ of rank $r$, while $\mathbb{H}$ is the maximal compact subalgebra of $\mathbb{G}$. One can then define the non-compact rank $r_{\mathrm{nc}}$ of $\mathcal{G}/\mathcal{H}$ as the dimension of the non-compact Cartan subalgebra $\mathrm{CSA}^{\mathrm{nc}}$:
\begin{equation} \label{defrank}
r_{\mathrm{nc}} \equiv \mathrm{dim}\ \mathrm{CSA}^{\mathrm{nc}} \,, \qquad \mathrm{CSA}^{\mathrm{nc}} \equiv \mathrm{CSA}_{\mathbb{G}_\mathbb{C}} \bigcap \mathbb{K} \,.
\end{equation}
When $r_{\mathrm{nc}} < r$, the manifold is called non-maximally non-compact, while for a so-called maximally non-compact space one has $r_{\mathrm{nc}} = r$. Note that in supergravity, the symmetric spaces that appear as target spaces of non-linear sigma models are generically non-maximally non-compact.

In the following section, we will consider cosmological solutions of 
 three-dimensional theories exhibiting symmetric, special quaternionic-K\"ahler geometry. After that, we will show how one can use the previous formulae to obtain time-dependent solutions of $d=4$, $\mathcal{N}=2$ supergravity.

\subsection{Step 2 : Obtaining cosmological solutions of $d=3$ supergravity with 8 supercharges} \label{ssec:obtsol}

The three-dimensional Lagrangian (\ref{3dlagrdual}) can be schematically written as a non-linear sigma model, coupled to gravity:
\begin{equation} \label{3dsigmamodel}
e^{-1}\tilde{\mathcal{L}}_{3d}  =  \frac{1}{2} R - \frac{1}{2} g_{AB}(\phi) \partial_\mu \phi^A \partial^\mu \phi^B \,,
\end{equation}
where we have collectively denoted the scalars by $\phi^A$. We will look for time-dependent solutions of this model, where the metric has a three-dimensional Friedmann-Lema\^{i}tre-Robertson-Walker form:
\begin{equation} \label{3dFLRW}
\rmd s^2_{3d} = -B^4(t)\, \rmd t^2 + B^2(t)\, \big( \rmd x^2 + \rmd y^2 \big) \,,
\end{equation}
where $B(t)$ is a function that needs to be determined by solving the Einstein equations.
In the background of a metric of the form (\ref{3dFLRW}), the equations of motion for the scalar fields are given by the geodesic equations in the target space with metric $g_{AB}(\phi)$:
\begin{equation} \label{3dgeodeq}
\frac{\rmd^2}{\rmd t^2} \phi^A + \Gamma^A_{BC} \frac{\rmd}{\rmd t} \phi^B \frac{\rmd}{\rmd t} \phi^C = 0 \,.
\end{equation}
It furthermore turns out that the function $B(t)$ in (\ref{3dFLRW}) is determined by the constant arclength $\gamma$ of the geodesic, traced out by the scalars:
\begin{equation} \label{solB}
B(t) = \exp \Big[\frac{\gamma}{\sqrt{2}} t \Big] \,, \qquad \gamma = \sqrt{g_{AB} \dot{\phi}^A \dot{\phi}^B} \,.
\end{equation}

The above discussion shows that a large class of cosmological solutions of the three-dimensional theory (\ref{3dsigmamodel}) is determined by the solutions of the geodesic equations (\ref{3dgeodeq}). In a series of papers \cite{Fre:2003ep,Fre:2005bs,Fre':2007hd} methods to solve these geodesic equations were developed for symmetric target spaces and as a consequence, it was shown that in this case the equations (\ref{3dgeodeq}) are integrable. In the following, we will shortly summarize the algorithmic approach to solve the geodesic equations (\ref{3dgeodeq}).

A crucial ingredient in the construction of the algorithm is a theorem \cite{Alekseevsky1975} that states that the symmetric space $\mathcal{G}/\mathcal{H}$ is metrically equivalent to a solvable group manifold $\mathcal{M}_{\mathrm{Solv}}$, obtained by exponentiating a solvable Lie algebra $\mathrm{Solv}(\mathcal{G}/\mathcal{H})$:
\begin{equation} \label{Misexpsolv}
\mathcal{M}_{\mathrm{Solv}} \equiv \exp \big[\mathrm{Solv}(\mathcal{G}/\mathcal{H}) \big] \,.
\end{equation}
For more details regarding the construction of this solvable Lie algebra, we refer to \cite{Andrianopoli:1996bq,Andrianopoli:1996zg,Fre:2001jd}. Furthermore, using a general theorem (see e.g. \cite{Helgason}), one can take a matrix representation of $\mathrm{Solv}(\mathcal{G}/\mathcal{H})$, such that all of its elements are given by upper triangular matrices. As a consequence, one can choose a coset representative $\mathbb{L}(\phi)$ that is given by the matrix exponential of an upper triangular matrix. We will comment on the precise definition of $\mathbb{L}(\phi)$, that is relevant in oxidizing the three-dimensional solutions to four dimensions, in the next section.

For purely time-dependent solutions, the coset representative $\mathbb{L}$ also becomes purely time-dependent, via its dependence on the supergravity scalars $\phi(t)$ : $\mathbb{L}(\phi(t)) = \mathbb{L}(t)$. Denoting by $\{\mathrm{K}_i\}$ and $\{\mathrm{H}_{\ell}\}$ orthonormal bases for $\mathbb{K}$ and $\mathbb{H}$ respectively, one defines the Lax operator $L(t)$ and the connection operator $W(t)$ as follows:
\begin{eqnarray} \label{defLaxW}
 L(t) & = & \sum_{i} \, \mbox{Tr} \left(\mathbb{L}^{-1}
  \frac{d}{dt}\mathbb{L} \, \mathrm{K}_i\right) \mathrm{K}_i \,,
  \nonumber\\
    W(t) & = & \sum_{\ell} \, \mbox{Tr} \left(\mathbb{L}^{-1}
  \frac{d}{dt}\mathbb{L} \, \mathrm{H}_\ell\right)\,
  \mathrm{H}_\ell \,.
\end{eqnarray}
Note that these are operators that depend on the first order time derivatives of the scalars. In \cite{Fre:2005bs}, it was shown that the geodesic equations (\ref{3dgeodeq}) for the coset manifold $\mathcal{G}/\mathcal{H}$ reduce to the matrix valued Lax equation:
\begin{equation} \label{Laxeq}
\frac{\rmd}{\rmd t} L = \big[L, W \big] \,.
\end{equation}
The connection $W(t)$ is moreover related to the Lax operator $L(t)$ as follows:
\begin{equation} \label{relWL}
W = L_{>0} - L_{<0} \,,
\end{equation}
where $L_{>0\ (<0)}$ denotes the strictly upper (resp. lower) triangular part of the Lax operator $L$. The relation (\ref{relWL}) stems from the fact that the coset representative $\mathbb{L}(\phi)$ from which the Lax operator is extracted is taken in the solvable parametrization. 

The Lax equation (\ref{Laxeq}) is a differential equation for the Lax operator $L$. Once an explicit solution $L_{\mathrm{sol}}(t)$ for $L(t)$ has been found, one is still left to solve a system of first order equations in order to find the solutions for the scalars fields. The latter first order system is obtained by comparing the definition of the Lax operator $L(t)$ in (\ref{defLaxW}) with the explicit solution $L_{\mathrm{sol}}(t)$. It turns out that, due to the choice of solvable parametrization, this last integration step can be performed in an iterative manner. We will refer to solving the Lax equation (\ref{Laxeq}) as 'the first integration step', while solving the resulting system of first order equations will be called 'the second integration step'.

The crucial step thus lies in solving the Lax equation (\ref{Laxeq}), with $W$ of the form (\ref{relWL}). This first integration step can however be performed in an algorithmic manner, as was shown in \cite{kodama2-1995,kodama-1995} and reviewed in \cite{Fre:2005bs,Fre':2007hd}. Essentially, the algorithm, which is nothing else
but an instance of the inverse scattering method, proceeds as follows. The equation (\ref{Laxeq}) represents the compatibility condition for the
following linear system exhibiting the iso-spectral property of $L$:
\begin{eqnarray}
\label{LaxIs}
L\Psi=\Psi \Lambda\,,\nonumber\\
\frac{\rmd}{\rmd t} \Psi=P\Psi\,,
\end{eqnarray}
where $\Psi(t)$ is the matrix whose $i$-th row is the
eigenvector $\varphi(t,\lambda_i)$ corresponding to the eigenvalue $\lambda_i$ of the Lax operator
$L(t)$ at time $t$ and $\Lambda$ is the diagonal
matrix of eigenvalues, which are constant throughout the whole time flow:
\begin{eqnarray}
\Psi&=&\left[\varphi(\lambda_1),\dots,\varphi(\lambda_n)]\equiv[\varphi_i(\lambda_j)\right]_{1\leq
i,j\leq n}\,,\nonumber\\
\Psi^{-1}&=&\left[\psi(\lambda_1),\dots,\psi(\lambda_n)]^T\equiv[\psi_j(\lambda_i)\right]_{1\leq
i,j\leq n}\,,\nonumber\\
 \Lambda&=& \mathrm{diag}\left(\lambda_1,\dots, \lambda_n \right)\,.
 \end{eqnarray}
The solution of (\ref{LaxIs}) for the Lax operator  is given by  the following explicit  form
of its matrix elements:
\begin{eqnarray}
\label{sol}\left[ L(t) \right]_{ij}=\sum_{k=1}^n \lambda_k
\varphi_i(\lambda_k,t) \psi_j(\lambda_k,t)~.
\end{eqnarray}
The eigenvectors of the Lax operator at each instant of time, which define the eigenmatrix $\Psi(t)$, and the
columns of its inverse $\Psi^{-1}(t)$, can be expressed in closed form in terms of
the initial data at some conventional instant of time, say at
$t=0$. Explicitly we have:
\begin{eqnarray} \label{Psi-1}
\varphi_i(\lambda_j,t)&=&\frac{\rme^{-\lambda_j
t}}{\sqrt{D_i(t)D_{i-1}(t)}} \, \mathrm{Det} \, \left ( \begin{array}{cccc}
c_{11}&\dots &c_{1,i-1}& \varphi_1^0(\lambda_j)\\
\vdots&\ddots&\vdots&\vdots\\
c_{i1}&\dots &c_{i,i-1}& \varphi_i^0(\lambda_j)\\
\end{array}\right
)\,,\nonumber\\
\psi_j(\lambda_i,t)&=&\frac{\rme^{-\lambda_i t}}{\sqrt{D_j(t)D_{j-1}(t)}}
\, \mathrm{Det} \, \left ( \begin{array}{ccc}
c_{11}&\dots &c_{1,j}\\
\vdots&\ddots&\vdots\\
c_{j-1,1}&\dots &c_{j-1,j}\\
 \psi_1^0(\lambda_i)&\dots &\psi_j^0(\lambda_i)
\end{array}\right)\,,
\end{eqnarray}
where the time-dependent matrix $c_{ij}(t)$ is defined as
\begin{eqnarray}
c_{ij}(t)=\sum_{k=1}^N \rme^{-2\lambda_k t}
\varphi_i^0(\lambda_k)\psi_j^0(\lambda_k)
\end{eqnarray}
and
\begin{eqnarray} \label{eigenvecat0}
  \varphi_i^0(\lambda_k) & := & \varphi_i(\lambda_k,0)~, \nonumber\\
  \psi_i^0(\lambda_k) & := &\psi_i(\lambda_k,0)
\end{eqnarray}
are the eigenvectors and their adjoints calculated at $t=0$.
These constant vectors as well as the eigenvalues $\lambda_k$ constitute the initial data of the
problem and provide the integration constants for the first integration step. Finally $D_k(t)$ denotes the determinant of the $k\times k$ matrix
with entries $c_{ij}(t)$:
 \begin{eqnarray} \label{Ddefi}
 D_k(t)=\mathrm{Det} \Biggr [ \Bigr ( c_{ij}(t) \Bigr )_{1\leq i,j \leq k} \Biggr
 ].
\end{eqnarray}
Note that $c_{ij}(0)=\delta_{ij}$ and $D_k(0)=1$.

Let us finally comment on how the initial conditions can be conveniently parametrized in this formalism. Note that at any instant of time $t$, the Lax operator $L(t)$ is an element of the Lie algebra $\mathbb{G}$ that lies in the orthogonal complement $\mathbb{K}$. Diagonalizing $L(t)$ as in (\ref{LaxIs}) then simply means that one brings the Lax operator inside the non-compact Cartan subalgebra $\mathrm{CSA}^{\mathrm{nc}}$. This can always be performed by conjugation with an element of the maximal compact subgroup $\mathcal{H}$. There thus exists a matrix $\mathcal{O} \in \mathcal{H}$, such that the Lax operator $L_0$ at $t=0$ can be written as
\begin{equation} \label{diagL0}
L_0 = \mathcal{O}^T\, \mathcal{C}_0\, \mathcal{O} \,,
\end{equation}
where $\mathcal{C}_0 \in \mathrm{CSA}^{\mathrm{nc}} $. The initial data can therefore be given as a pair
\begin{equation} \label{initialdata}
\mathcal{C}_0 \in \mathrm{CSA} \bigcap \mathbb{K} \,, \qquad \mathcal{O} \in \mathcal{H} \,.
\end{equation}
The matrix $\mathcal{O}$ then neatly summarizes the initial conditions (\ref{eigenvecat0}), while $\mathcal{C}_0$ contains the eigenvalues $\lambda_k$ of the Lax operator.

\subsection{Step 3 : Uplifting to four-dimensional solutions} \label{ssec:uplift}

We will now explain how the time-dependent solutions of the model described by (\ref{3dlagrdual}) can be uplifted to solutions of the four-dimensional theory (\ref{4dlagr}). The problem consists in finding a coset representative $\mathbb{L}(\phi)$ such that the coset metric, derived from it, leads to a non-linear sigma model of the form displayed in (\ref{3dlagrdual}). 

The key point in constructing the correct coset representative is the observation that the three-dimensional U-duality algebra $\mathbb{G} = \mathbb{U}_{d=3}$ admits the following decomposition \cite{Fre':2005si}:
\begin{equation} \label{decomp3d4dalg}
\mathrm{adj}(\mathbb{U}_{d=3}) = \mathrm{adj}(\mathbb{U}_{d=4}) \oplus \mathrm{adj}(\SL(2,\mathbb{R})_E) \oplus W_{(\mathbf{W},\mathbf{2})} \,.
\end{equation}
In the above decomposition, $\mathbb{U}_{d=4}$ represents the U-duality algebra in four dimensions. $W$ contains generators that transform in the representation $(\mathbf{W},\mathbf{2})$ with respect to the adjoint action of $\mathrm{adj}(\mathbb{U}_{d=4}) \oplus \mathrm{adj}(\SL(2,\mathbb{R})_E)$, where $\mathbf{W}$ denotes the symplectic representation of $\mathbb{U}_{d=4}$. As a consequence of this decomposition, the solvable algebra that represents the homogeneous quaternionic-K\"ahler manifold, determining the sigma model in three dimensions, admits the following decomposition:
\begin{equation} \label{decomp3d4dsolvalg}
\mathrm{adj}(\mathrm{Solv}(\mathbb{U}_{d=3})) = \mathrm{adj}(\mathrm{Solv}(\mathbb{U}_{d=4})) \oplus \mathrm{adj}(\mathrm{Solv}(\SL(2,\mathbb{R})_E)) \oplus W_{(\mathbf{W},\mathbf{1_+})}\,,
\end{equation}
where by the notation $W_{(\mathbf{W},\mathbf{1_+})}$ we indicate that we take half of the generators contained in $W_{(\mathbf{W},\mathbf{2})}$, namely the generators that have positive grading with respect to $\SL(2,\mathbb{R})_E$. $\mathrm{Solv}(\mathbb{U}_{d=4})$ then denotes the solvable algebra that represents the homogeneous special K\"ahler manifold, that determines the sigma model spanned by the four-dimensional scalar fields. 

In order to discuss the above decompositions in more detail, we will resort to the description of homogeneous quaternionic-K\"ahler manifolds that was devised by Alekseevsky. It was shown in \cite{Alekseevsky1975,deWit:1991nm,deWit:1993wf,Cortes} that homogeneous quaternionic-K\"ahler manifolds are metrically equivalent to solvable group manifolds, whose solvable algebras have a particular structure and are correspondingly called quaternionic algebras. More specifically, the ranks of these quaternionic algebras can range from 1 to 4. In case the rank is equal to 4, the algebra is characterized by three integers $q$, $P$ and $\dot{P}$ and its generators can be summarized in the following scheme:
\begin{equation}
 \label{weightsVgeneral}
\begin{array}{|llll|l|}
\hline
h_0\,  &g_0\,:\,(1,0,0,0) & q_0\,:\,(\frac{1}{2},-\frac{1}{2},-\frac{1}{2},-\frac{1}{2}) & p_0\,:\,(\frac{1}{2},\frac{1}{2},\frac{1}{2},\frac{1}{2}) & 1 \\
h_1\, & g_1\,:\,(0,1,0,0) & q_1\,:\,(\frac{1}{2},-\frac{1}{2},\frac{1}{2},\frac{1}{2}) & p_1\,:\,(\frac{1}{2},\frac{1}{2},-\frac{1}{2},-\frac{1}{2}) & 1 \\
h_2\, & g_2\,:\,(0,0,1,0) & q_2\,:\,(\frac{1}{2},\frac{1}{2},-\frac{1}{2},\frac{1}{2}) & p_2\,:\,(\frac{1}{2},-\frac{1}{2},\frac{1}{2},-\frac{1}{2}) & 1 \\
 h_3\, & g_3\,:\,(0,0,0,1) & q_3\,:\,(\frac{1}{2},\frac{1}{2},\frac{1}{2},-\frac{1}{2}) & p_3\,:\,(\frac{1}{2},-\frac{1}{2},-\frac{1}{2},\frac{1}{2}) & 1 \\
 X^+\,:\,(0,0,\frac{1}{2},\frac{1}{2}) & X^-\,:\,(0,0,\frac{1}{2},- \frac{1}{2}) & \tilde X^+\,:\,(\frac{1}{2},\frac{1}{2},0,0) & \tilde X^-\,:\,(\frac{1}{2},- \frac{1}{2},0,0) & q \\
Y^+\,:\,(0,\frac{1}{2},0,\frac{1}{2}) & Y^-\,:\,(0,\frac{1}{2},0,- \frac{1}{2}) & \tilde Y^+\,:\,(\frac{1}{2},0,\frac{1}{2},0) & \tilde Y^-\,:\,(\frac{1}{2},0,-\frac{1}{2},0) & \multirow{2}{*}{$(P + \dot{P})\mathcal{D}_{q+1}$} \\
Z^+\,:\,(0,\frac{1}{2},\frac{1}{2},0) & Z^-\,:\,(0,\frac{1}{2},-\frac{1}{2},0) & \tilde Z^+\,:\,(\frac{1}{2},0,0,\frac{1}{2}) & \tilde Z^-\,:\,(\frac{1}{2},0,0,-\frac{1}{2}) &  \\
\hline
\end{array}
\end{equation}
Each entry in the above diagram represents either one generator or a multiplet of generators. All entries in one row of the diagram however contain an equal amount of generators, that is given in the last column of the table. Note that $\mathcal{D}_{q+1}$ denotes the dimension of the irreducible representations of the real Clifford algebra in $q+1$ Euclidean dimensions. These dimensions can for instance be found in \cite{deWit:1991nm}. The generators in an entry of type $Y$, together with the generators in the entry of type $Z$ below it, span a space of dimension $(P + \dot{P}) \mathcal{D}_{q+1}$. The dimension of the manifold $\mathcal{M}_Q$ obtained by exponentiating the quaternionic algebra represented by (\ref{weightsVgeneral}) is thus given by:
\begin{equation} \label{dimMQ}
\mathrm{dim}\ \mathcal{M}_Q = 4 (n_V + 1) \,, \qquad n_V = 3 + q + (P + \dot{P}) \mathcal{D}_{q+1} \,.
\end{equation}
Note that $n_V$ is indeed equal to the number of vector multiplets that was present in the four-dimensional supergravity that leads to the quaternionic algebra (\ref{weightsVgeneral}) upon dimensional reduction. 

The Cartan subalgebra of the quaternionic algebra is given by $\{h_0, h_1, h_2, h_3\}$. The gradings of the other generators in the diagram (\ref{weightsVgeneral}) with respect to this Cartan subalgebra are indicated in brackets. When an entry in the diagram represents a multiplet of generators, all generators in this multiplet have the same gradings. In case the rank of the quaternionic algebra is less than four, a similar scheme  can be obtained by making a suitable truncation of (\ref{weightsVgeneral}). We refer to \cite{Fre:2006eu} for more details regarding this point.

With reference to the decomposition (\ref{decomp3d4dsolvalg}), we see that the following identifications hold:
\begin{equation}
\renewcommand\arraystretch{1.2}
\begin{array}{|c|c|}
\hline \mathrm{Component\  of}\ \mathrm{adj}(\mathrm{Solv}(\mathbb{U}_{d=3})) & \mathrm{generators} \\ \hline
\hline \mathrm{adj}(\mathrm{Solv}(\SL(2,\mathbb{R})_E)) & h_0, g_0 \\ \hline \mathrm{adj}(\mathrm{Solv}(\mathbb{U}_{d=4})) & h_1, h_2, h_3, g_1, g_2, g_3, X^+, X^-, Y^+, Y^-, Z^+, Z^- \\ \hline W_{(\mathbf{W},\mathbf{1_+})}  & q_0, q_1, q_2, q_3, p_0, p_1, p_2, p_3, \tilde{X}^+, \tilde{X}^-, \tilde{Y}^+, \tilde{Y}^-, \tilde{Z}^+, \tilde{Z}^- \\ \hline
\end{array}
\end{equation}
Note that the generators identified with $W_{(\mathbf{W},\mathbf{1_+})}$ indeed all have positive grading with respect to the Cartan generator $h_0$ of $\SL(2,\mathbb{R})_E$.

From a physical point of view, each generator in the table (\ref{weightsVgeneral}) is associated to a scalar field in the three-dimensional supergravity. Referring to (\ref{reducscheme}), the generators identified with $\mathrm{adj}(\mathrm{Solv}(\mathbb{U}_{d=4}))$ correspond to the scalars $w^\alpha$, that were already present in four dimensions. The generators in $\mathrm{adj}(\mathrm{Solv}(\SL(2,\mathbb{R})_E))$ correspond to the scalars $\Delta$ and $\omega$, obtained from the reduction of the metric, while the generators of $W_{(\mathbf{W},\mathbf{1_+})}$ are associated to the scalars $\chi^I$ and $\sigma_I$, that come from reducing the four-dimensional vectors $\hat{A}^I_{\hat{\mu}}$. The fact that the generators of $W$ form a symplectic representation of the four-dimensional U-duality group is then easily understood as a consequence of electric-magnetic duality in four dimensions \cite{Gaillard:1981rj}. Indeed, in four dimensions, electric-magnetic duality acts as an invariance of the combined system of equations of motion and Bianchi identities. Under these duality transformations, the field strengths $\hat{\mathcal{F}}^I_{\hat{\mu} \hat{\nu}}$, together with their duals $\hat{\mathcal{G}}_{I \hat{\mu} \hat{\nu}}$ transform as a symplectic vector under the U-duality group:
\begin{equation}
\left(\begin{array}{c} \hat{\mathcal{F}}^I \\ \hat{\mathcal{G}}_I \end{array} \right) \ \ \ \longrightarrow \ \ \ \mathcal{S} \left(\begin{array}{c} \hat{\mathcal{F}}^I \\ \hat{\mathcal{G}}_I \end{array} \right) \,, \qquad \mathcal{S} \in \Sp(2 n_V + 2, \mathbb{R}) \,,
\end{equation}
where the dual field strengths $\hat{\mathcal{G}}_{I \hat{\mu} \hat{\nu}}$ are defined as 
\begin{equation} \label{defdualfieldstr}
\hat{\mathcal{G}}_{I \hat{\mu} \hat{\nu}}= \rmi \varepsilon_{\hat{\mu} \hat{\nu}\hat{\rho}\hat{\sigma}} \frac{\delta \hat{\mathcal{L}}_{4d}}{\delta \hat{\mathcal{F}}^I_{\hat{\rho} \hat{\sigma}}} \,.
\end{equation}
Since upon dimensional reduction the scalars $\chi^I$ come from the reduction of the vectors via (\ref{redvectors}), while the $\sigma_I$ come from the duals of the vectors via (\ref{dualityrels}), it is therefore not surprising that their associated generators in $W$ also transform as a symplectic vector under the four-dimensional U-duality group.

Using the above discussion, we can now present the precise formula for the coset representative $\mathbb{L}(\phi)$. Our notations concerning the scalar fields associated to the different generators are summarized in the following table:
\begin{equation} \label{schemegenscalars}
\begin{array}{|c|c|}
\hline \mathrm{generator} & \mathrm{scalar} \\ \hline 
h_0 & \log(\Delta(t)) \\ 
g_0 & \omega(t) \\
h_k\,, \ \ k=1, \ldots,3 & \rmh^k(t)\,, \ \ k=1, \ldots,3 \\
g_k\,, \ \ k=1, \ldots,3 & \rmg^k(t)\,, \ \ k=1, \ldots,3 \\
X^\pm_r\,, \ \ r=1,\ldots,q &  \rmX^{\pm\, r}(t)\,, \ \ r=1,\ldots,q \\
Y^\pm_s\,, \ \ s=1,\ldots,(P+\dot{P}) \mathcal{D}_{q+1}/2 &  \rmY^{\pm\, s}(t)\,, \ \ s=1,\ldots,(P+\dot{P}) \mathcal{D}_{q+1}/2\\
Z^\pm_s\,, \ \ s=1,\ldots,(P+\dot{P}) \mathcal{D}_{q+1}/2 &  \rmZ^{\pm\, s}(t)\,, \ \ s=1,\ldots,(P+\dot{P}) \mathcal{D}_{q+1}/2 \\
p_i\,, \ \ i=0,\ldots,3 & \rmp^i(t)\,, \ \ i=0,\ldots,3 \\
q_i\,, \ \ i=0,\ldots,3 & \rmq^i(t)\,, \ \ i=0,\ldots,3 \\ 
\tilde{X}^\pm_r\,, \ \ r=1,\ldots,q &  \tilde{\rmX}^{\pm\, r}(t)\,, \ \ r=1,\ldots,q \\
\tilde{Y}^\pm_s\,, \ \ s=1,\ldots,(P+\dot{P}) \mathcal{D}_{q+1}/2 &  \tilde{\rmY}^{\pm\, s}(t)\,, \ \ s=1,\ldots,(P+\dot{P}) \mathcal{D}_{q+1}/2\\
\tilde{Z}^\pm_s\,, \ \ s=1,\ldots,(P+\dot{P}) \mathcal{D}_{q+1}/2 &  \tilde{\rmZ}^{\pm\, s}(t)\,, \ \ s=1,\ldots,(P+\dot{P}) \mathcal{D}_{q+1}/2 \\ \hline
\end{array}
\end{equation}
With the above notations, the coset representative is calculated as follows:
\begin{equation} \label{calccosetrepr}
\mathbb{L}(t) = \exp\big[\omega(t) g_0\big] \ \rme^W \ \mathbb{L}_{SK}\ \exp\Big[\log\big(\Delta(t)\big) h_0\Big] \,,
\end{equation}
where $\mathbb{L}_{SK}$ contains the scalar fields that were already present in four dimensions:
\begin{eqnarray} \label{defLSK}
\mathbb{L}_{SK} & = & \rme^{\rmX^{-\, r}(t) X^-_r} \ \rme^{\rmX^{+\, r}(t) X^+_r} \ \rme^{\rmg^3(t) g_3}\ \rme^{\rmY^{-\, s}(t) Y^-_s}  \ \rme^{\rmY^{+\, s}(t) Y^+_s}  \nonumber \\ & & Ê\quad \rme^{\rmg^2(t) g_2} \ \rme^{\rmZ^{-\, t}(t) Z^-_t} \ \rme^{\rmZ^{+\, t}(t) Z^+_t} \ \rme^{\rmg^1(t) g_1}\ \rme^{\sum_{k=1}^3 \rmh^k(t) h_k}\,,
\end{eqnarray}
and $\rme^W$ contains the scalars of type $\chi^I$ and $\sigma_I$:
\begin{eqnarray} \label{defeW}
\rme^W & = & \exp\Big[\sqrt{2}\Big(\rmp^i(t) p_i + \rmq^i(t) q_i + \tilde{\rmX}^{+\, r}(t) \tilde{X}^+_r + \tilde{\rmY}^{+\, s}(t) \tilde{Y}^+_s + \tilde{\rmZ}^{+\, t}(t) \tilde{Z}^+_t\nonumber \\ & & \qquad \qquad + \tilde{\rmX}^{-\, r}(t) \tilde{X}^-_r + \tilde{\rmY}^{-\, s}(t) \tilde{Y}^-_s + \tilde{\rmZ}^{-\, t}(t) \tilde{Z}^-_t\Big) \Big] \,.
\end{eqnarray}
Which of the scalars in $\rme^W$ are of type $\chi^I$ and which are of type $\sigma_I$ can be determined group theoretically, by studying the symplectic representation $\mathbf{W}$ of the four-dimensional U-duality group, in which these scalars transform. This will be made more clear in a specific example in the next section.

Using the coset representative $\mathbb{L}$ in (\ref{calccosetrepr}) one can calculate the metric on the coset space in the usual way. With the order of exponentiation of (\ref{calccosetrepr}) and the identifications for the scalar fields made above, this metric indeed has the structure displayed in (\ref{3dlagrdual}). In particular, it is straightforward to extract the period matrix $\mathcal{N}_{IJ}$ from the expression of the coset metric. This shows that this identification of the coset representative $\mathbb{L}$ and the scalars is the one that allows to easily uplift three-dimensional solutions to four dimensions. Once a solution for the three-dimensional scalar fields has been found, one simply has to use the various formulae of section \ref{ssec:red} to obtain the solutions for the four-dimensional fields. 

In order to reconstruct the four-dimensional metric and vector fields, one has to use the formulas (\ref{redmetric}) and (\ref{redvectors}), that contain the vector fields that were present in three dimensions before dualization to scalar fields. These can be obtained by integrating the equations (\ref{dualityrels}). This integration is however particularly simple when taking into account the Bianchi identities for the three-dimensional vectors and the ansatz for the three-dimensional solutions, explained in section \ref{ssec:obtsol}. Indeed, by using the ansatz (\ref{3dFLRW}) for the three-dimensional metric, one obtains that the only non-zero components of the three-dimensional field strengths are:
\begin{eqnarray} \label{nonzerocompfieldstr}
\mathcal{F}^{(0)}_{xy} & = &  \frac{1}{\Delta^2} \left[ \partial_t \omega + \chi^I \stackrel{\leftrightarrow}{\partial_t} \sigma_I \right] \,, \nonumber \\
\F^{I}_{xy} & = & - \frac{1}{\Delta^2} \bigg[ \Delta \left(\I \N \right)^{-1|IJ} \left( (\R \N)_{JK} \partial_t \chi^K - \partial_t \sigma_J \right) \nonumber \\Ê& & +\  \chi^I \left( \partial_t \omega + \chi^J \stackrel{\leftrightarrow}{\partial_t} \sigma_J \right) \bigg] \,.
\end{eqnarray}
The Bianchi identities for the three-dimensional field strenghts $\mathcal{F}^{(0)}_{\mu \nu}$, $\F^{I}_{\mu \nu}$ then imply that
\begin{equation} \label{bianchi3d}
\partial_t \mathcal{F}^{(0)}_{xy} = 0 \,, \qquad \partial_t \F^{I}_{xy} = 0 \,.
\end{equation}
In other words, $\mathcal{F}^{(0)}_{xy}$ and $\F^{I}_{xy}$ are constant for the solutions under consideration. Note that this gives a non-trivial check for the solutions for the three-dimensional scalar fields, generated by the Lax algorithm, as it implies that the right-hand sides of (\ref{nonzerocompfieldstr}) should be constants of motion of the flow. Denoting the constant values of $\mathcal{F}^{(0)}_{xy}$, $\F^{I}_{xy}$ by $\alpha$, $\beta^I$ respectively, a possible solution for the vectors in three dimensions is then given by
\begin{eqnarray} \label{solvectors3d}
(A^{(0)}_t, A^{(0)}_x, A^{(0)}_y) & = & \Big(0, -\frac{\alpha}{2} y, \frac{\alpha}{2} x\Big) \,, \nonumber \\
(A^{I}_t, A^{I}_x, A^{I}_y) & = & \Big(0, -\frac{\beta^I}{2} y, \frac{\beta^I}{2} x\Big) \,.
\end{eqnarray}
Using these formulas, full four-dimensional solutions can be easily constructed. The four-dimensional metric is for instance explicitly given by:
\begin{equation}
\rmd s^2_{4d} = -\frac{B^4(t)}{\Delta(t)} \rmd t^2 + \frac{B^2(t)}{\Delta(t)}\left(\rmd x^2 + \rmd y^2\right) + \Delta(t) \left(\rmd z + A^{(0)}_\mu \rmd x^\mu \right)^2 \,,
\end{equation}
with $B(t)$ given by (\ref{solB}) and $A^{(0)}_\mu$ determined above. One can also see that the four-dimensional electric and magnetic fields only have the following non-zero components:
\begin{eqnarray} \label{nonzerocompfieldstr4d}
\hat{\F}^I_{t x} & = & (\partial_t \chi^I) A^{(0)}_x \,, \nonumber \\
\hat{\F}^I_{t y} & = & (\partial_t \chi^I) A^{(0)}_y \,, \nonumber \\
\hat{\F}^I_{t z} & = & \partial_t \chi^I \,, \nonumber \\
\hat{\F}^I_{xy} & = & \mathcal{F}^I_{xy} + \chi^I \mathcal{F}^{(0)}_{xy} = \beta^I + \alpha \chi^I \,.
\end{eqnarray}
Finally, note that due to the fact that according to (\ref{dimMQ}) the dimension of the three-dimensional sigma model is equal to $4(n_V + 1)$, the four-dimensional solutions will depend on twice this number of integration constants.

In the following section, we will consider a specific $d=4$, $\mathcal{N}=2$ supergravity model. We will use the above described method to find time-dependent solutions and discuss some of their properties.

\section{Examples} \label{sec:examples}

\subsection{The choice of model} \label{ssec:choiceex}

In order to give some examples of four-dimensional cosmologies obtained with the described method, we will consider a three-dimensional supergravity for which the non-linear sigma model (\ref{3dlagrdual}) is determined by the following symmetric quaternionic-K\"ahler manifold of rank 4:
\begin{equation} \label{MQisso44}
\mathcal{M}_Q = \frac{\SO(4,4)}{\SO(4) \times \SO(4)} \,.
\end{equation}
This model can be uplifted to $d=4$, $\mathcal{N}=2$ supergravity, coupled to three vector multiplets. The scalar fields in four dimensions then span the following special K\"ahler manifold of complex dimension three:
\begin{equation} \label{MSKisSL23}
\mathcal{M}_{SK} = \frac{\SL(2,\mathbb{R})}{\SO(2)} \times \frac{\SO(2,2)}{\SO(2) \times \SO(2)} = \frac{\SL(2,\mathbb{R})}{\SO(2)} \times \frac{\SL(2,\mathbb{R})}{\SO(2)} \times \frac{\SL(2,\mathbb{R})}{\SO(2)} \,.
\end{equation}
The choice of this model is determined both by the desire to have a model that is computationally easy and yet captures interesting behavior, as by the fact that the above model has also a nice ten-dimensional interpretation. Indeed, as we will discuss in more detail in section \ref{sec:10d}, this case corresponds to the well-known $S$-$T$-$U$-model, that can be obtained by performing a compactification of type IIB supergravity on a $K3\times T^2/\mathbb{Z}_2$-orientifold.

In order to discuss the relation between the solvable coordinates (\ref{schemegenscalars}) and the scalars $w^\alpha$, $\chi^I$, $\sigma_I$, $\Delta$, $\omega$ that appear in (\ref{3dlagrdual}), we will start from the decompositions (\ref{decomp3d4dalg}) and (\ref{decomp3d4dsolvalg}), which for our example are given by:
\begin{eqnarray} \label{decomps3d4dalgappl}
\mathrm{adj}\big(\SO(4,4)\big) & = &\mathrm{adj}\big(\SL(2,\mathbb{R})^3\big) \oplus \mathrm{adj}\big(\SL(2,\mathbb{R})_E\big) \oplus W_{(\mathbf{2},\mathbf{2},\mathbf{2},\mathbf{2})} \,, \nonumber \\
\mathrm{adj}\Big(\mathrm{Solv}\big(\SO(4,4)\big)\Big) & = & \mathrm{adj}\Big(\mathrm{Solv}\big(\SL(2,\mathbb{R})^3\big)\Big) \oplus \mathrm{adj}\Big(\mathrm{Solv}\big(\SL(2,\mathbb{R})_E\big)\Big) \oplus W_{(\mathbf{2},\mathbf{2},\mathbf{2},\mathbf{1_+})} \,.
\end{eqnarray}
The solvable algebra that generates the symmetric space (\ref{MQisso44}) is a truncation of the algebra given in (\ref{weightsVgeneral}):
\begin{equation}
 \label{weightsSO44}
\begin{array}{|llll|}
\hline
 h_0\,:\,(0,0,0,0)  &g_0\,:\,(1,0,0,0) & q_0\,:\,(\frac{1}{2},-\frac{1}{2},-\frac{1}{2},-\frac{1}{2}) & p_0\,:\,(\frac{1}{2},\frac{1}{2},\frac{1}{2},\frac{1}{2}) \\
 h_1\,:\,(0,0,0,0)  & g_1\,:\,(0,1,0,0) & q_1\,:\,(\frac{1}{2},-\frac{1}{2},\frac{1}{2},\frac{1}{2}) & p_1\,:\,(\frac{1}{2},\frac{1}{2},-\frac{1}{2},-\frac{1}{2}) \\
 h_2\,:\,(0,0,0,0)  & g_2\,:\,(0,0,1,0) & q_2\,:\,(\frac{1}{2},\frac{1}{2},-\frac{1}{2},\frac{1}{2}) & p_2\,:\,(\frac{1}{2},-\frac{1}{2},\frac{1}{2},-\frac{1}{2}) \\
 h_3\,:\,(0,0,0,0)  & g_3\,:\,(0,0,0,1) & q_3\,:\,(\frac{1}{2},\frac{1}{2},\frac{1}{2},-\frac{1}{2}) & p_3\,:\,(\frac{1}{2},-\frac{1}{2},-\frac{1}{2},\frac{1}{2}) \\
\hline
\end{array}
\end{equation}
In comparison with (\ref{weightsVgeneral}) there are no generators of type $X$, $Y$ or $Z$. All generators mentioned in the above diagram are therefore non-degenerate. The above diagram then makes the decomposition (\ref{decomps3d4dalgappl}) more explicit, since $\{h_0, g_0\}$ are identified as the Cartan generator and positive root of $\SL(2,\mathbb{R})_E$, while $h_i$, $g_i$, $i=1,2,3$ correspond to a Cartan generator and positive root of the three $\SL(2,\mathbb{R})$-factors that constitute the four-dimensional U-duality group. The generators of type $p$ and $q$ that have positive grading with respect to $h_0$, span the representation $W$. From the gradings one infers that these indeed transform in the symplectic representation $\mathbf{W}=(\mathbf{2},\mathbf{2},\mathbf{2})$ of $\SL(2,\mathbb{R})^3$.

The coset representative $\mathbb{L}$ is then constructed as in (\ref{calccosetrepr},\ref{defLSK},\ref{defeW}) upon ignoring all factors of type $X$, $Y$ or $Z$. With reference to the notations of (\ref{schemegenscalars}), the scalar fields $\mathrm{h}^k(t)$, $\mathrm{g}^k(t)$, $k=1,2,3$ can be combined in the three complex scalars that parametrize the special K\"ahler manifold (\ref{MSKisSL23}). The scalars $\mathrm{p}^i(t)$, $\mathrm{q}^i(t)$ can be identified with the fields $\chi^I$, $\sigma_I$ that appear upon dimensional reduction of the four vector fields. We will determine which of the $\mathrm{p}^i(t)$, $\mathrm{q}^i(t)$ are "electric" (i.e. of type $\chi^I$) and which are "magnetic" (i.e. of type $\sigma_I$) by using their grading with respect to $h_1$. The generators with positive grading with respect to $h_1$ are then associated to the $\chi^I$-scalars, while the ones with negative grading are assigned to the scalars of type $\sigma_I$. Furthermore, our identification is such that the scalars of type $\chi^I$ and their corresponding $\sigma_I$ have opposite gradings with respect to $h_1$, $h_2$ and $h_3$. In summary, we obtain the following identification:
\begin{equation} \label{identchisigmapq}
\renewcommand\arraystretch{1.2}
\begin{array}{|ccc|ccc|} \hline \chi^0 & \leftrightarrow & \mathrm{p}^0(t) & \sigma_0 & \leftrightarrow & \mathrm{q}^0(t) \\ \hline \chi^1 & \leftrightarrow & \mathrm{q}^3(t) & \sigma_1 & \leftrightarrow & \mathrm{p}^3(t) \\ \hline \chi^2 & \leftrightarrow & \mathrm{q}^2(t) & \sigma_2 & \leftrightarrow & \mathrm{p}^2(t) \\ \hline \chi^3 & \leftrightarrow & \mathrm{p}^1(t) & \sigma_3 & \leftrightarrow & \mathrm{q}^1(t) \\ \hline \end{array}
\end{equation}
Using these identifications, the metric on the coset space can be calculated. The components have the form dictated by (\ref{3dlagrdual}). Using the Lax algorithm, described in section \ref{ssec:obtsol}, explicit time-dependent solutions for the fields of the three-dimensional supergravity can be determined by use of a computer program. By virtue of the above identifications, these solutions can then be interpreted as solutions of the four-dimensional theory, as explained in section \ref{ssec:uplift}. Note that, from the expression for the metric on the coset space (\ref{MQisso44}), one can infer the period matrix that determines the couplings of the scalar fields to the vector fields in the four-dimensional Lagrangian (\ref{4dlagr}). The explicit expression for this matrix is given in appendix \ref{app:periodmatrix}. Appendix B then contains the conventions we used in constructing the solvable algebra of $\mathcal{M}_Q$.

In the following, we will give some explicit examples of four-dimensional cosmological solutions obtained by our method. The initial conditions, needed as input for the Lax algorithm, will be parametrized as in (\ref{initialdata}). In this case, the diagonal matrix $\mathcal{C}_0$ is given by:
\begin{equation} \label{defci}
\mathcal{C}_0 = c^i h_i \ \ \ (i=0,\ldots,3) \,,
\end{equation}
and is thus determined by giving the four constants $c^i$. The matrix $\mathcal{O}$ is an element of $\SO(4) \times \SO(4)$. In the following, we will parametrize it using 12 Euler angles. Denoting the 12 positive roots of the $\SO(4,4)$ algebra (corresponding to $g_i$, $p_i$, $q_i$, $(i=0,\ldots,3)$) collectively by $E^{\alpha_I}$, $(I=1,\ldots,12)$, one writes:
\begin{equation} \label{Eulerangles}
\mathcal{O} = \prod_{I=1}^{12} \exp\Big[\theta_I \big(E^{\alpha_I} - E^{\alpha_I T}\big)\Big] \,.
\end{equation}
We refer to appendix \ref{app:solvalgsoconv} for more information regarding our conventions for the solvable algebra of $\SO(4,4)$ and the precise identification of the $E^{\alpha_I}$ generators. The 16 initial conditions for the Lax algorithm are then given by the 4 constants $c^i$ and the 12 Euler angles $\theta_I$. Note that the second integration step introduces another set of 16 integration constants. The solutions will thus in general depend on 32 integration constants.

\subsection{Asymptotic behavior of the solutions} \label{ssec:asympsol}

Before turning to explicit solutions, let us first comment on their general behavior. The asymptotic behavior of the solutions can be inferred from the asymptotic behavior of the Lax operator. It was shown in \cite{kodama2-1995,kodama-1995} that the Lax operator becomes constant and diagonal at asymptotic times $t = \pm \infty$. Furthermore it was noted in \cite{Fre':2007hd} that these asymptotic states are connected to the state of the solution at $t=0$ via elements of the Weyl group $W(\SO(4,4))$ of $\mathcal{G}=\SO(4,4)$. Due to (\ref{diagL0}), (\ref{defci}) and the conventions summarized in appendix \ref{app:solvalgsoconv}, the Lax operator at $t=0$ can be diagonalized to 
\begin{equation} \label{C0expl}
\mathcal{C}_0 = \mathrm{diag}\Big(\frac{c^0+c^1}{2},\frac{c^0-c^1}{2},\frac{c^2+c^3}{2},\frac{c^2-c^3}{2},-\frac{c^2-c^3}{2},-\frac{c^2+c^3}{2},-\frac{c^0-c^1}{2},-\frac{c^0+c^1}{2}\Big)\,.
\end{equation}
As explained in more detail in appendix \ref{app:weylso44}, the Weyl group $W(\SO(4,4))$ naturally acts on the constants $c^i$ in a linear fashion:
\begin{equation} \label{actweylgroup}
\sigma \in W \quad : \quad c^i \qquad \longmapsto \qquad \sigma(c^i) \,.
\end{equation}
The Lax operators $L_{\pm \infty} = \lim_{t \rightarrow \pm \infty}\ L(t)$ are then related to $\mathcal{C}_0$ via the action of two elements $\sigma_{\pm \infty} \in W(\SO(4,4))$:
\begin{eqnarray} \label{asLax}
L_{-\infty} & = & \mathrm{diag}\Big(\sigma_{-\infty}\Big(\frac{c^0+c^1}{2}\Big),\sigma_{-\infty}\Big(\frac{c^0-c^1}{2}\Big),\sigma_{-\infty}\Big(\frac{c^2+c^3}{2}\Big),\sigma_{-\infty}\Big(\frac{c^2-c^3}{2}\Big),\nonumber \\ &  & \quad -\sigma_{-\infty}\Big(\frac{c^2-c^3}{2}\Big),-\sigma_{-\infty}\Big(\frac{c^2+c^3}{2}\Big),-\sigma_{-\infty}\Big(\frac{c^0-c^1}{2}\Big),-\sigma_{-\infty}\Big(\frac{c^0+c^1}{2}\Big)\Big)\nonumber \,,\\
L_{+\infty} & = & \mathrm{diag}\Big(\sigma_{+\infty}\Big(\frac{c^0+c^1}{2}\Big),\sigma_{+\infty}\Big(\frac{c^0-c^1}{2}\Big),\sigma_{+\infty}\Big(\frac{c^2+c^3}{2}\Big),\sigma_{+\infty}\Big(\frac{c^2-c^3}{2}\Big),\nonumber\\ &  & \quad -\sigma_{+\infty}\Big(\frac{c^2-c^3}{2}\Big),-\sigma_{+\infty}\Big(\frac{c^2+c^3}{2}\Big),-\sigma_{+\infty}\Big(\frac{c^0-c^1}{2}\Big),-\sigma_{+\infty}\Big(\frac{c^0+c^1}{2}\Big)\Big) \,.
\end{eqnarray}
Let us introduce the following notation for the Cartan scalars at asymptotic times:
\begin{eqnarray}
\Delta_{\pm \infty}(t) & \equiv & \lim_{t\rightarrow \pm \infty} \Delta(t) \,, \nonumber \\
\mathrm{h}^k_{\pm \infty}(t) & \equiv & \lim_{t\rightarrow \pm \infty} \mathrm{h}^k(t) \,,  \quad k=1,2,3 \,.
\end{eqnarray}
The system of differential equations to be solved in the second integration step, then reduces for asymptotic times $t = \pm \infty$ to:
\begin{eqnarray} \label{asdiffeqs}
\frac{\dot{\Delta}_{\pm \infty}(t)}{\Delta_{\pm \infty}(t)} & = & \sigma_{\pm \infty}(c^0) \,, \nonumber \\
\dot{\mathrm{h}}^k_{\pm \infty}(t) & = & \sigma_{\pm \infty}(c^k) \,, \quad k=1,2,3 \,, \nonumber \\
\dot{\mathrm{g}}^i(t) & = & \dot{\mathrm{p}}^i(t) \ = \ \dot{\mathrm{q}}^i(t)\ = \ 0 \,. \quad i = 0, \cdots, 3 \,.
\end{eqnarray}
We thus find that at asymptotic times, the solution simplifies to:
\begin{equation} \label{assolutions}
\renewcommand\arraystretch{1.3}
\begin{array}{|c|c|}
\hline \mathrm{For}\ t=-\infty & \mathrm{For}\ t= + \infty \\ \hline
\log{\Delta}_{-\infty}(t) =  \sigma_{-\infty}(c^0) t + a^0_{-\infty}& \log \Delta_{+\infty}(t) = \sigma_{+\infty}(c^0) t + a^0_{+\infty} \\ \hline \mathrm{h}^1_{-\infty}(t) = \sigma_{-\infty}(c^1) t + a^1_{-\infty} & \mathrm{h}^1_{+\infty}(t) = \sigma_{+\infty}(c^1) t + a^1_{+\infty} \\ \hline \mathrm{h}^2_{-\infty}(t) = \sigma_{-\infty}(c^2) t + a^2_{-\infty} & \mathrm{h}^2_{+\infty}(t) = \sigma_{+\infty}(c^2) t + a^2_{+\infty} \\ \hline \mathrm{h}^3_{-\infty}(t) = \sigma_{-\infty}(c^3) t + a^3_{-\infty} & \mathrm{h}^3_{+\infty}(t) = \sigma_{+\infty}(c^3) t + a^3_{+\infty} \\ \hline 
\end{array}
\end{equation}
where $a^i_{\pm \infty}$, $i = 0, \cdots, 3$ are integration constants introduced in the second integration step. The other scalars are just constant at asymptotic times. We thus see that the non-trivial time-dependence of the solutions at asymptotic times is determined by the integration constants $c^i$, that serve as input for the Lax algorithm and by the Weyl group elements $\sigma_{\pm \infty}$. Stated differently, to each solution, one can associate a unique Weyl group element $\sigma = \sigma_{+\infty} \sigma_{-\infty}^{-1}$ that acts on asymptotic states of the form in (\ref{assolutions}) as:
\begin{eqnarray} \label{actweyl}
\sigma \in W(\SO(4,4)) & : &   \frac{\dot{\Delta}_{-\infty}(t)}{\Delta_{-\infty}(t)} = \sigma_{-\infty}(c^0) \quad  \mapsto \quad  \sigma\Big(\frac{\dot{\Delta}_{-\infty}(t)}{\Delta_{-\infty}(t)}\Big)  =  \frac{\dot{\Delta}_{+\infty}(t)}{\Delta_{+\infty}(t)} = \sigma_{+\infty}(c^0) \,,\nonumber \\
& & \ \,  \dot{\mathrm{h}}^k_{-\infty}(t) = \sigma_{-\infty}(c^k) \ \ \, \, \mapsto \quad \sigma\Big(\dot{\mathrm{h}}^k_{-\infty}(t)\Big) = \dot{\mathrm{h}}^k_{+\infty}(t) = \sigma_{+\infty}(c^k) \,.
\end{eqnarray}
The solutions produced by the Lax algorithm thus interpolate between two asymptotic states of the form given in (\ref{assolutions}) according to the action of the Weyl group element $\sigma$.

So far, we have discussed the asymptotics of our time-dependent solutions and the corresponding action of the Weyl group in terms of the three-dimensional scalar fields. Now we can also interpret the above discussion in terms of asymptotics of four-dimensional fields. Let us first of all note that the fact that only the three-dimensional Cartan fields exhibit non-trivial time-dependent behavior at asymptotic times, can be confirmed by considering the behavior of the different terms in (\ref{4dlagr}) for explicit solutions. As can be confirmed in explicit examples, the terms involving the vector fields in the four-dimensional Lagrangian, rapidly tend to zero when $t\rightarrow \pm \infty$. The only non-trivial asymptotic dynamics is thus contained in the first two terms of (\ref{4dlagr}). Moreover, regarding the second term, involving the scalar fields, inspection of our solutions shows that only the Cartan fields $\mathrm{h}^k(t)$, $k=1,2,3$ evolve non-trivially at asymptotic times. As the four-dimensional sigma-model (\ref{MSKisSL23}) is endowed with a Euclidean metric when truncated to the Cartan fields, we can truncate the action (\ref{4dlagr}) to the following one:
\begin{eqnarray} \label{4dlagrrestr}
\hat{\mathcal{L}}_{4d} & = & \frac{1}{2} \hat{e}\hat{R} - \frac{1}{4} \hat{e}\delta_{i j} \partial_{\hat{\mu}} \mathrm{h}^i \partial^{\hat{\mu}} \mathrm{h}^{j} \,,
\end{eqnarray}
for the purpose of investigating the asymptotic behavior of the solutions.
Adopting the following ansatz for the four-dimensional metric at $t=\pm \infty$:
\begin{equation} \label{asmetric4d}
\rmd s_{\pm \infty}^2 = -\frac{B_{\pm \infty}(t)^4}{\Delta_{\pm \infty}(t)} \rmd t^2 + \frac{B_{\pm \infty}(t)^2}{\Delta_{\pm \infty}(t)}\big(\rmd x^2 + \rmd y^2\big) + \Delta_{\pm \infty}(t) \rmd z^2 \,,
\end{equation}
one can obtain the following equations of motion from the Lagrangian (\ref{4dlagrrestr}):
\begin{eqnarray} \label{eqsrestr}
\ddot{\mathrm{h}}_{\pm \infty}^i & = & 0 \,, \quad i=1,2,3\,, \nonumber \\
\dot{\Delta}_{\pm \infty}^2(t) & = & \Delta_{\pm \infty}(t) \ddot{\Delta}_{\pm \infty}(t) \,, \nonumber \\
4 \frac{\dot{B}_{\pm \infty}(t)^2}{B_{\pm \infty}(t)^2} & = & \dot{\mathrm{h}}_{\pm \infty}^i \dot{\mathrm{h}}_{\pm \infty}^i + \frac{\dot{\Delta}_{\pm \infty}(t)^2}{\Delta_{\pm \infty}(t)^2} \,.
\end{eqnarray}
The first two of these equations imply that at $t=-\infty$ and $t=+\infty$, the solutions for $\Delta(t)$ and $\mathrm{h}^i(t)$ are indeed of the form given in (\ref{assolutions}). Using these solutions in the Einstein equations, the functions $B_{\pm \infty}(t)$ are determined by the following equation:
\begin{equation} \label{diffBpminfty}
\frac{\dot{B}_{\pm \infty}(t)^2}{B_{\pm \infty}(t)^2} = \frac{1}{4} \sum_{k=0}^3 \left[\sigma_{\pm \infty}(c^k)\right]^2 \,.
\end{equation}
Note however that according to (\ref{orthpropactweylc}), the quantities on the right-hand-side of this equation are invariant under the Weyl group, i.e.:
\begin{equation} \label{defalpha}
\sum_{k=0}^3 \left[\sigma_{- \infty}(c^k)\right]^2 = \sum_{k=0}^3 \left[\sigma_{+ \infty}(c^k)\right]^2 \equiv \alpha^2 \,.
\end{equation}
This implies that the functions $B_{\pm \infty}(t)$ obey the same differential equation. 
This equation is solved by an exponential function:
\begin{equation} \label{Bpminfty}
B_{\pm \infty}(t) = C_{\pm \infty} \exp\left[\frac{\alpha}{2} t\right] \,,
\end{equation}
where $C_{\pm \infty}$ are integration constants. By comparing with the form of the exact solution (\ref{solB}), we see that
\begin{equation} \label{propBpminfty}
C_{- \infty} = C_{+ \infty} \,, \qquad \gamma = \frac{\alpha}{\sqrt{2}} \,, \qquad B_{-\infty}(t) = B_{+ \infty}(t) = B(t) \,.
\end{equation}
Using the above consideration, the action of the Weyl group element $\sigma = \sigma_{+\infty} \sigma_{-\infty}^{-1}$ can then be interpreted as changing the metric at $t=-\infty$:
\begin{equation} \label{metricmininfty}
\rmd s^2 = -\exp\big[(2 \alpha - \sigma_{-\infty}(c^0))t\big] \rmd t^2 + \exp\big[(\alpha - \sigma_{-\infty}(c^0))t\big]\big( \rmd x^2 + \rmd y^2 \big) + \exp\big[\sigma_{-\infty}(c^0)t\big] \rmd z^2 \,,
\end{equation}
to the following metric at $t=+\infty$\footnote{Here and in the previous formula, we have absorbed some trivial integration constants in the definition of the coordinates $t$, $x$, $y$ and $z$.}:
\begin{equation} \label{metricplusinfty}
\rmd s^2 = -\exp\big[(2 \alpha - \sigma_{+\infty}(c^0))t\big] \rmd t^2 + \exp\big[(\alpha - \sigma_{+\infty}(c^0))t\big]\big( \rmd x^2 + \rmd y^2 \big) + \exp\big[\sigma_{+\infty}(c^0)t\big] \rmd z^2 \,.
\end{equation}
The action of the Weyl group on the four-dimensional solution thus consists in changing the scale factors of the asymptotic form of the metric as well as changing the behavior of the scalar fields $\mathrm{h}^1(t)$, $\mathrm{h}^2(t)$, $\mathrm{h}^3(t)$ as indicated in (\ref{assolutions}).

\subsection{Three examples} \label{ssec:threeex}

Let us now illustrate the above discussion with three examples. The examples are given in order of increasing complexity. For each of the three examples, we will take initial conditions for which the eigenvalues of the Lax operator at $t=0$ are determined by the following values of $c^i$ in (\ref{defci}):
\begin{equation} \label{ciex}
c^0 = 1 \,, \quad c^1 = 2 \,, \quad c^2 = 3 \,,  \quad c^4 = 4 \,.
\end{equation}
The initial conditions that correspond to the Euler angles $\theta_I$ in (\ref{Eulerangles}) will however be taken different for the different examples. We will each time explicitly mention which choice for these angles was taken.

In order to give a physical interpretation of the solutions, it is also useful to analyze the various parts of the four-dimensional energy-momentum tensor. The full energy-momentum tensor derived from (\ref{4dlagr}) consists of two parts:
\begin{equation} \label{emtensor}
T^{tot}_{\hat{\mu} \hat{\nu}} = T^0_{\hat{\mu} \hat{\nu}} + T^1_{\hat{\mu} \hat{\nu}} \,,
\end{equation}
where $T^0_{\mu \nu}$ is associated to the four-dimensional scalars fields:
\begin{equation} \label{emtensor0}
T^0_{\hat{\mu} \hat{\nu}} = 2 g_{\alpha \bar{\beta}} \partial_{\hat{\mu}} w^\alpha \partial_{\hat{\nu}} \bar{w}^{\bar{\beta}} - g_{\alpha \bar{\beta}} \partial_{\hat{\rho}} w^\alpha \partial_{\hat{\sigma}} \bar{w}^{\bar{\beta}} \hat{g}^{\hat{\rho} \hat{\sigma}} \hat{g}_{\hat{\mu} \hat{\nu}} \,,
\end{equation}
and $T^1_{\mu \nu}$ denotes the contribution of the vector fields:
\begin{equation} \label{emtensor1}
T^1_{\hat{\mu} \hat{\nu}} = \frac{1}{4} \big(\mathrm{Im} \mathcal{N}_{IJ} \big)\hat{\F}^I_{\hat{\rho} \hat{\sigma}} \hat{\F}^{J \hat{\rho} \hat{\sigma}} \hat{g}_{\hat{\mu} \hat{\nu}} - \big(\mathrm{Im} \mathcal{N}_{IJ}\big) \hat{\F}^I_{\hat{\mu} \hat{\rho}} \hat{\F}^J_{\hat{\nu} \hat{\sigma}} \hat{g}^{\hat{\rho} \hat{\sigma}} \,.
\end{equation}
Note that in the examples, the energy-momentum tensor and metric will often (but not always) be diagonal. We will then define the energy and pressure densities associated to the full energy-momentum tensor or the part associated to the scalars or the vectors respectively in the following way:
\begin{eqnarray}
T^{tot, 0, 1}_{00} & = & - g_{00} \, \rho^{tot, 0, 1} \,, \nonumber \\
T^{tot, 0, 1}_{ii} & = &  g_{ii} \, P_i^{tot, 0, 1} \,. 
\end{eqnarray}

\subsubsection{Example 1} \label{sssec:ex1}

Our first example is characterized by the following choice for the Euler angles (\ref{Eulerangles}):
\begin{equation} \label{anglesex1}
\theta_1 = \frac{\pi}{3} \,, \qquad \theta_I = 0\ \ \ (I=2,\ldots,12) \,.
\end{equation}
Applying the Lax algorithm with these initial conditions as input, leads to the following solutions for the three-dimensional scalar fields:
{\renewcommand{\arraystretch}{1.3}
\begin{longtable}{|rcl|rcl|} 
\hline
$\omega(t)$  &  $=$  & $C[1]\,,$ & $\mathrm{p}^0(t)$  & $=$ & $C[8]\,,$ \\
$\Delta(t)$ & $=$ & $\rme^t C[9]\,,$ & $\mathrm{p}^1(t)$ & $=$ & $C[13]\,,$ \\
$\mathrm{h}^1(t)$  & $=$ & $2 t+C[2]-\log\left[1+3 \rme^{4 t}\right]\,,$ & $\mathrm{p}^2(t)$ & $=$ & $C[14]\,,$  \\
$\mathrm{h}^2(t)$ &  $=$  & $3 t+C[6]\,,$ & $\mathrm{p}^3(t)$ & $=$ & $C[15]\,,$ \\
$\mathrm{h}^3(t)$  & $=$ & $4 t+C[7]\,,$  & $\mathrm{q}^0(t)$ & $=$ & $C[16]\,,$  \\
$\mathrm{g}^1(t)$ & $=$ & $\displaystyle{-\frac{\rme^{C[2]}}{\sqrt{3} \left(1+3 \rme^{4 t}\right)}}+C[3]\,,$ & $\mathrm{q}^1(t)$ & $=$ & $C[10]\,,$  \\
$\mathrm{g}^2(t)$ & $=$  & $C[4]\,,$  & $\mathrm{q}^2(t)$ & $=$ & $C[11]\,,$  \\
$\mathrm{g}^3(t)$ &  $=$ & $C[5]\,,$ & $\mathrm{q}^3(t)$ & $=$ & $C[12]\,.$  \\ \hline
\caption{The solutions for the three-dimensional scalars with $c^0=1$, $c^1=2$, $c^3=3$, $c^4=4$ and $\theta_1=\pi/3$.} \label{sol1}
\end{longtable}} \vspace{-5mm}
Note that the above solution not only depends on the initial conditions (\ref{ciex}) and (\ref{anglesex1})  for the Lax integration, but also on 16 integration constants $C[1],\ldots,C[16]$ that appear in the second integration step, as explained in section \ref{ssec:obtsol}. One can moreover explicitly check that the right-hand sides of equations (\ref{nonzerocompfieldstr}) are constants of motion. In this specific case, the constants $\alpha$ and $\beta^I$ are zero. 

Using the solutions from table \ref{sol1}, one can construct a solution of the four-dimensional field equations. The four-dimensional metric is given by:
\begin{equation} \label{formmetricscalefactors}
\rmd s^2 = -r^{[0]}(t)^2 \rmd t^2 + r^{[1]}(t)^2 \big(\rmd x^2 + \rmd y^2\big) + r^{[2]}(t)^2 \rmd z^2 \,,
\end{equation}
where the various scale factors are given by:
\begin{eqnarray} \label{scalefacsol1}
r^{[0]}(t)^2 & = & \frac{\rme^{\left(-1+2 \sqrt{30}\right) t}}{C[9]} \,, \nonumber \\
r^{[1]}(t)^2 & = & \frac{\rme^{\left(-1+ \sqrt{30}\right) t}}{C[9]} \,, \nonumber \\
r^{[2]}(t)^2 & = & \rme^t C[9]\,. 
\end{eqnarray}
The solutions for the four-dimensional scalar fields $\mathrm{h}^1(t)$, $\mathrm{h}^2(t)$, $\mathrm{h}^3(t)$, $\mathrm{g}^1(t)$, $\mathrm{g}^2(t)$, $\mathrm{g}^3(t)$ can be found in table \ref{sol1}, while the four-dimensional field strengths $\hat{\F}^I_{\hat{\mu} \hat{\nu}}$ are all zero for this solution. Thus the energy momentum tensor gets a contribution coming only from the scalar fields. Explicitly, it is given by:
\begin{equation} \label{emsol1}
T^{tot}_{\hat{\mu} \hat{\nu}} =\left(
\begin{array}{cccc}
 \frac{29}{4} & 0 & 0 & 0 \\
 0 & \frac{29}{4} \rme^{-\sqrt{30} t} & 0 & 0 \\
 0 & 0 & \frac{29}{4} \rme^{-\sqrt{30} t} & 0 \\
 0 & 0 & 0 & \frac{29}{4} \rme^{-2 \left(-1+\sqrt{30}\right) t} C[9]^2
\end{array}
\right)\,.
\end{equation}
The energy and pressure densities in the various directions are then given by:
\begin{eqnarray} \label{rhoPsol1}
\rho & = & \frac{29}{4} \rme^{-\left(-1+2 \sqrt{30}\right) t} C[9] \,, \nonumber \\
P_1 & = & \frac{29}{4} \rme^{t-2 \sqrt{30} t} C[9] = P_2 = P_3 \,.
\end{eqnarray}

Note the behavior of the scalar $\mathrm{h}^1(t)$ in the above solution. For $t \rightarrow -\infty$, $\mathrm{h}^1(t)$ behaves as $2t$, while for $t \rightarrow + \infty$ this behavior is inverted to $-2t$. This behavior is caused by the action of the Weyl group as explained in section \ref{ssec:asympsol}. For this specific solution, $\sigma_{-\infty}$ acts as the identity on the constants $c^i$, whereas $\sigma_{+\infty}$ acts as:
\begin{equation} \label{sigmaplussol1}
\sigma_{+\infty}(c^0) = c^0\,, \quad \sigma_{+\infty}(c^1) = - c^1\,, \quad \sigma_{+\infty}(c^2) = c^2\,, \quad \sigma_{+\infty}(c^3) = c^3\,.
\end{equation}
According to (\ref{assolutions}), this implies that the asymptotic behavior of $\Delta(t)$, $\mathrm{h}^2(t)$, $\mathrm{h}^3(t)$ is not changed in going from $t=-\infty$ to $t=+\infty$, while the behavior of $\mathrm{h}^1(t)$ indeed changes with a minus sign.

Finally, let us give the solution in terms of the cosmic time $\tau$, by making the coordinate transformation:
\begin{eqnarray} \label{defcosmictime}
\rmd \tau & = & r^{[0]}(t) \rmd t \,, \nonumber \\
\tau & = & \int_{-\infty}^t r^{[0]}(t')\, \rmd t' \,.
\end{eqnarray}
In terms of this new time coordinate, the four-dimensional metric becomes:
\begin{equation} \label{formmetricscalefactorsnew}
\rmd s^2 = -\rmd \tau^2 + \tilde{r}^{[1]}(\tau)^2 \big(\rmd x^2 + \rmd y^2\big) + \tilde{r}^{[2]}(\tau)^2 \rmd z^2 \,,
\end{equation}
where
\begin{eqnarray} \label{scalefactausol1}
\tilde{r}^{[1]}(\tau)^2 & = &\frac{\left(-\frac{1}{2}+\sqrt{30}\right)^{\frac{2 \left(-1+\sqrt{30}\right)}{-1+2 \sqrt{30}}} \left(\tau \sqrt{C[9]}\right)^{\frac{2 \left(-1+\sqrt{30}\right)}{-1+2 \sqrt{30}}}}{C[9]}\,, \nonumber \\
\tilde{r}^{[2]}(\tau)^2 & = & \left(-\frac{1}{2}+\sqrt{30}\right)^{\frac{2}{-1+2 \sqrt{30}}} \left(\tau \sqrt{C[9]}\right)^{\frac{2}{-1+2 \sqrt{30}}} C[9]\,. 
\end{eqnarray}
This allows us to determine the time evolution of the total volume $V(\tau) = (\tilde{r}^{[1]}(\tau))^2 \tilde{r}^{[2]}(\tau)$ of the four-dimensional space-time:
\begin{equation} \label{totvolsol1}
V(\tau) = \left(-\frac{1}{2}+\sqrt{30}\right) \tau \,.
\end{equation}
We thus find that the total volume grows linearly with cosmic time $\tau$. In terms of the cosmic time, the energy and pressure densities are given by:
\begin{eqnarray} \label{rhoPtausol1}
\rho & = & \frac{29}{\left(1-2 \sqrt{30}\right)^2} \frac{1}{\tau^2}\,, \nonumber \\
P_1 & = & \frac{29}{\left(1-2 \sqrt{30}\right)^2} \frac{1}{\tau^2} \ = \ P_2 \ = \ P_3 \,.
\end{eqnarray}

\subsubsection{Example 2} \label{sssec:ex2}

For our second example, we choose the following values for the Euler angles (\ref{Eulerangles}):
\begin{equation} \label{anglesex2}
\theta_2 = \frac{\pi}{3} \,, \qquad \theta_I = 0\ \ \ (I \neq 2) \,.
\end{equation}
This leads to the following solutions for the three-dimensional scalar fields:
{\renewcommand{\arraystretch}{1.8}
\begin{longtable}{|rcl|} 
\hline \multicolumn{3}{r}{Continued on next page}
\endfoot
\hline
\endhead
\endlastfoot
\hline & &  \\[-7mm]
$\omega(t)$  &  $=$  & $\displaystyle{\frac{1}{2} \bigg(\frac{\sqrt{6} \rme^{\frac{1}{2} (C[3]-C[4]-C[5])} (-C[1] C[7]-C[2] (C[6]+C[1] C[8])+C[9]) \sqrt{C[10]}}{3+\rme^{4 t}}}$ \\ & & $+2 C[11]\bigg)\,,$   \\
$\Delta(t)$ & $=$ & $\displaystyle{\frac{\rme^{3 t} C[10]}{\sqrt{3+\rme^{4 t}}}} \,, $ \\
$\mathrm{h}^1(t)$  & $=$ & $C[3]+2 \Big(2 t-\frac{1}{4} \log\big[3+\rme^{4 t}\big]\Big) \,,$ \\
$\mathrm{h}^2(t)$  & $=$ & $C[4]+3 \Big(\frac{t}{3}+\frac{1}{6} \log\big[3+\rme^{4 t}\big]\Big)\,,$ \\
$\mathrm{h}^3(t)$  & $=$ & $2 t+C[5]+\frac{1}{2} \log\big[3+\rme^{4 t}\big]\,,$ \\
$\mathrm{g}^1(t)$ & $=$ & $C[16]\,,$ \\
$\mathrm{g}^2(t)$ & $=$ & $C[1]\,,$ \\
$\mathrm{g}^3(t)$ & $=$ & $C[2]\,,$ \\
$\mathrm{p}^0(t)$  & $=$ & $\displaystyle{-\frac{\sqrt{\frac{3}{2}} \rme^{\frac{1}{2} (C[3]-C[4]-C[5])} C[1] C[2] \sqrt{C[10]}}{3+\rme^{4 t}}+C[12]}\,,$ \\
$\mathrm{p}^1(t)$  & $=$ & $\displaystyle{\frac{\sqrt{\frac{3}{2}} \rme^{\frac{1}{2} (C[3]-C[4]-C[5])} \sqrt{C[10]}}{3+\rme^{4 t}}+C[13]}\,,$ \\
$\mathrm{p}^2(t)$  & $=$ & $C[6]\,,$ \\
$\mathrm{p}^3(t)$  & $=$ & $C[7]\,,$ \\
$\mathrm{q}^0(t)$ & $=$ & $C[8] \,,$ \\
$\mathrm{q}^1(t)$ & $=$ & $C[9] \,,$ \\
$\mathrm{q}^2(t)$ & $=$ & $\displaystyle{-\frac{\sqrt{\frac{3}{2}} \rme^{\frac{1}{2} (C[3]-C[4]-C[5])} C[2] \sqrt{C[10]}}{3+\rme^{4 t}}+C[14]} \,,$ \\
$\mathrm{q}^3(t)$ & $=$ & $\displaystyle{\frac{\sqrt{\frac{3}{2}} \rme^{\frac{1}{2} (C[3]-C[4]-C[5])} C[1] \sqrt{C[10]}}{3+\rme^{4 t}}+C[15]} \,.$ \\[2mm]
\hline
\caption{The solutions for the three-dimensional scalars with $c^0=1$, $c^1=2$, $c^3=3$, $c^4=4$ and $\theta_2=\pi/3$.} \label{sol2}
\end{longtable}} \vspace{-5mm}
One can again explicitly check that this solution is such that $\mathcal{F}^{(0)}_{xy}$ and  $\mathcal{F}^{I}_{xy}$ are constant. More specifically, one obtains:
\begin{eqnarray} \label{alphabetasol2}
\alpha & = & 0 \,, \nonumber \\
\beta^0 & = & -\frac{2 \sqrt{6} \rme^{\frac{1}{2} (-C[3]+C[4]+C[5])} C[16]}{\sqrt{C[10]}} \,, \nonumber \\
\beta^1 & = & 0 \ =\  \beta^2\  =\  \beta^3\,.
\end{eqnarray}

Let us again look at the asymptotic behavior of the Cartan scalars, summarized in the following table:
\begin{equation} \label{asympcsasol2}
\renewcommand{\arraystretch}{1.2}
\begin{array}{|c|c|}
\hline t \rightarrow -\infty & t \rightarrow + \infty \\ \hline 
\log(\Delta(t)) \sim 3 t & \log(\Delta(t)) \sim t \\ 
\mathrm{h}^1(t)  \sim  4 t & \mathrm{h}^1(t)  \sim  2 t \\
\mathrm{h}^2(t)  \sim  t & \mathrm{h}^2(t)  \sim 3 t \\ 
\mathrm{h}^3(t)  \sim  2 t & \mathrm{h}^3(t)  \sim  4 t \\ \hline
\end{array}
\end{equation}
Again this behavior can be explained in terms of the action of the Weyl group on the initial conditions $c^i$. In this case, the Weyl group element $\sigma_{+ \infty}$ acts as the identity on the constants $c^i$, while $\sigma_{-\infty}$ acts as:
\begin{equation} \label{actsigmamininfex2}
\setlength{\arraycolsep}{5mm}
\begin{array}{ll}\displaystyle{\sigma_{-\infty}(c^0)\  =  \ \frac{c^0}{2} - \frac{c^1}{2} + \frac{c^2}{2} + \frac{c^3}{2}} \,, &
\displaystyle{\sigma_{-\infty}(c^1)\  = \ -\frac{c^0}{2} + \frac{c^1}{2} + \frac{c^2}{2} + \frac{c^3}{2}} \,, \\[2mm]
\displaystyle{\sigma_{-\infty}(c^2)\  = \  \frac{c^0}{2} + \frac{c^1}{2} + \frac{c^2}{2} - \frac{c^3}{2}} \,, & 
\displaystyle{\sigma_{-\infty}(c^3)\  = \ \frac{c^0}{2} + \frac{c^1}{2} - \frac{c^2}{2} + \frac{c^3}{2}} \,.
\end{array}
\end{equation}
Using the explicit values (\ref{ciex}), this action of the Weyl group indeed reconstructs the asymptotic behavior (\ref{asympcsasol2}), in agreement with the discussion in section \ref{ssec:asympsol}.

\begin{figure}[!t]
\begin{center}
\subfigure[Scale factor 1]{\label{plotsscalefactorssol2-a}\scalebox{0.80}{\includegraphics[width=7.5cm]{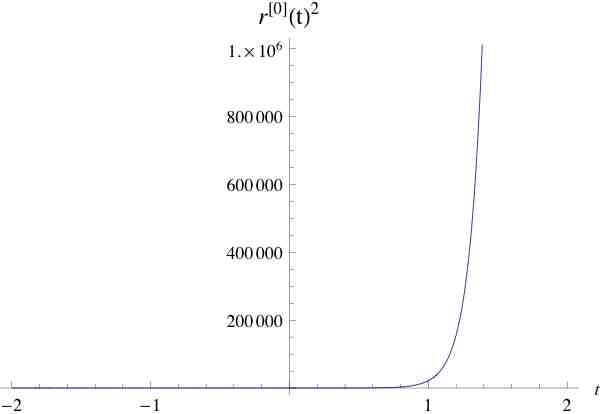}}} 
\subfigure[Scale factor 2]{\label{plotsscalefactorssol2-b}\scalebox{0.80}{\includegraphics[width=7.5cm]{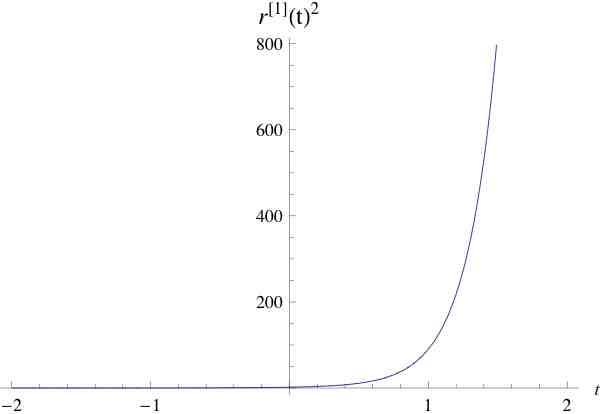}}} \\
\subfigure[Scale factor 3]{\label{plotsscalefactorssol2-c}\scalebox{0.80}{\includegraphics[width=7.5cm]{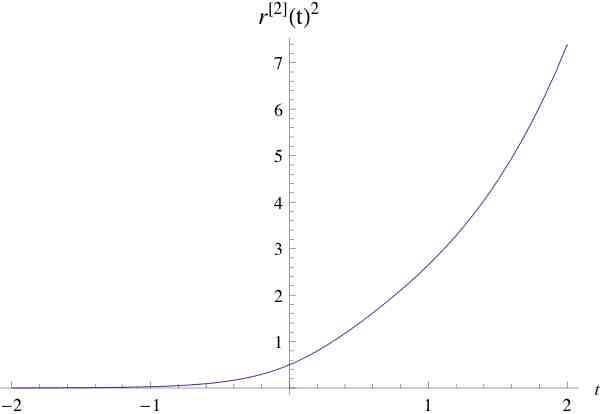}}}
\end{center}
\caption{The scale factors determining the metric of example 2.} 
\label{plotsscalefactorssol2}
\end{figure}

\begin{figure} [!t]
\begin{center}
\subfigure[Electric field 1]{\label{plotselfieldssol2-a}\scalebox{0.80}{\includegraphics[width=7.5cm]{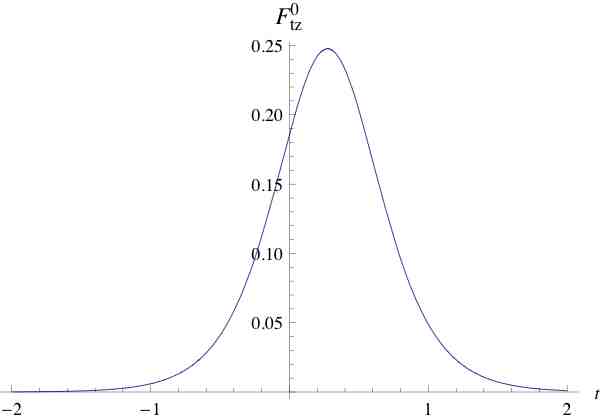}}} 
\subfigure[Electric field 2]{\label{plotselfieldssol2-b}\scalebox{0.80}{\includegraphics[width=7.5cm]{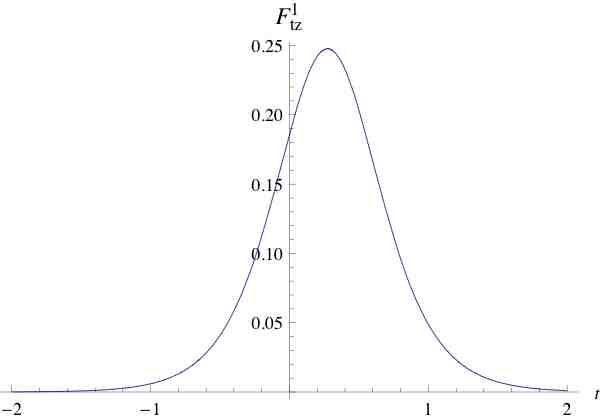}}} \\
\subfigure[Electric field 3]{\label{plotselfieldssol2-c}\scalebox{0.80}{\includegraphics[width=7.5cm]{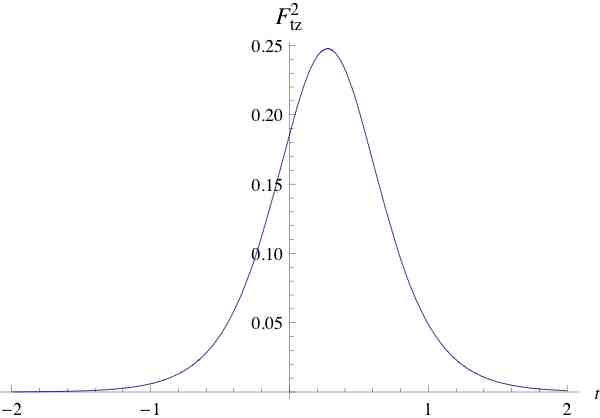}}} 
\subfigure[Electric field 4]{\label{plotselfieldssol2-d}\scalebox{0.80}{\includegraphics[width=7.5cm]{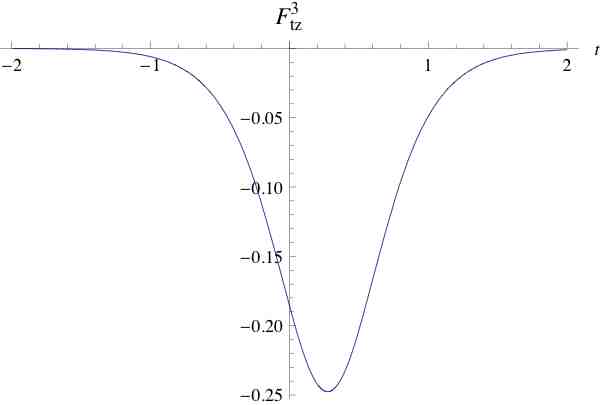}}}
\end{center}
\caption{Electric fields for the second solution.} 
\label{plotselfieldssol2}
\end{figure}

\begin{figure} [!t]
\begin{center}
\subfigure[Total energy density]{\label{emtotsol2-a}\scalebox{0.80}{\includegraphics[width=7.5cm]{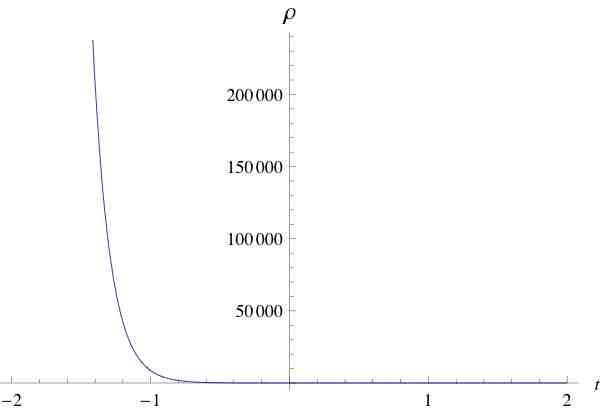}}} 
\subfigure[$P_1^{tot}$]{\label{emtotsol2-b}\scalebox{0.80}{\includegraphics[width=7.5cm]{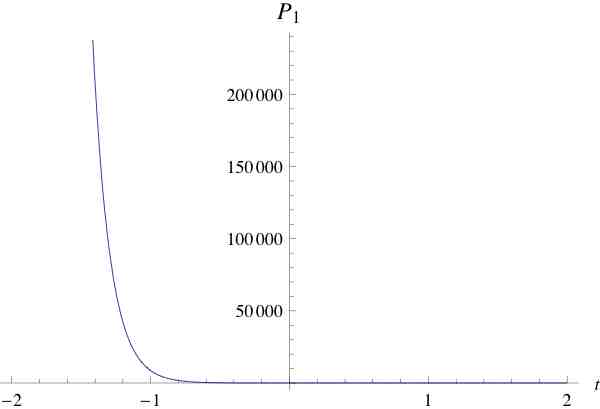}}} \\
\subfigure[$P_3^{tot}$]{\label{emtotsol2-c}\scalebox{0.80}{\includegraphics[width=7.5cm]{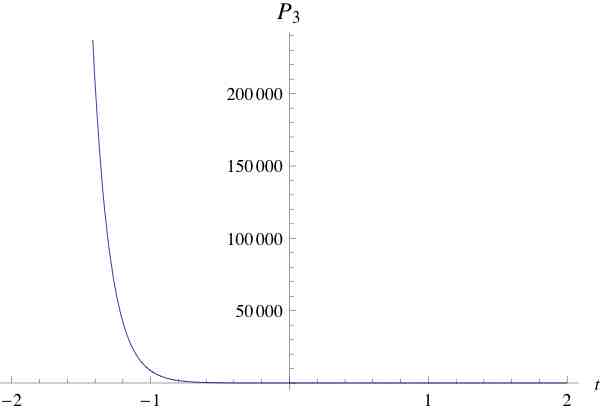}}}
\end{center}
\caption{Total energy density and pressure densities for the second solution.} 
\label{emtotsol2}
\end{figure}

\begin{figure} [!t]
\begin{center}
\subfigure[Scalar part of the energy density]{\label{emscalarsol2-a}\scalebox{0.70}{\includegraphics[width=7.5cm]{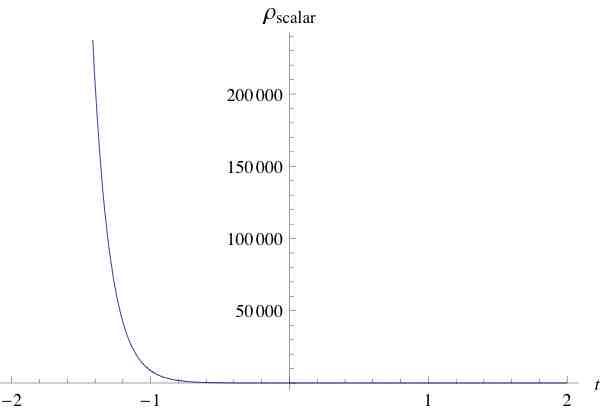}}} 
\subfigure[$P_1^{0}$]{\label{emscalarsol2-b}\scalebox{0.70}{\includegraphics[width=7.5cm]{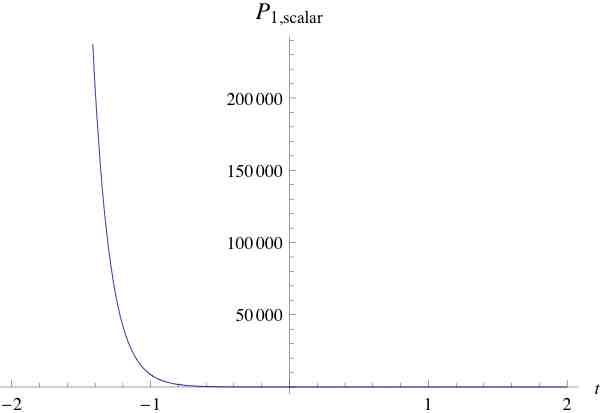}}} \\
\subfigure[$P_3^{0}$]{\label{emscalarsol2-c}\scalebox{0.70}{\includegraphics[width=7.5cm]{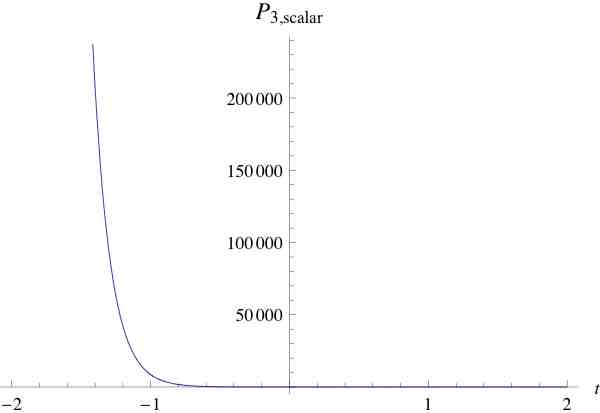}}} 
\end{center}
\caption{Energy density and pressure densities associated to the scalar fields for the second solution.} 
\label{emscalarsol2}
\end{figure}

\begin{figure} [!h]
\begin{center}
\subfigure[Vector part of the energy density]{\label{emvectorsol2-a}\scalebox{0.70}{\includegraphics[width=7.5cm]{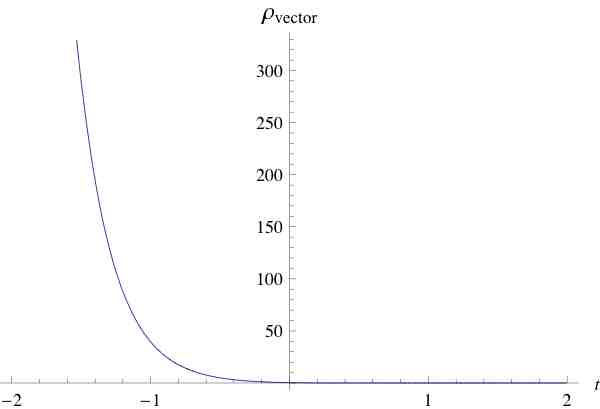}}} 
\subfigure[$P_1^{1}$]{\label{emvectorsol2-b}\scalebox{0.70}{\includegraphics[width=7.5cm]{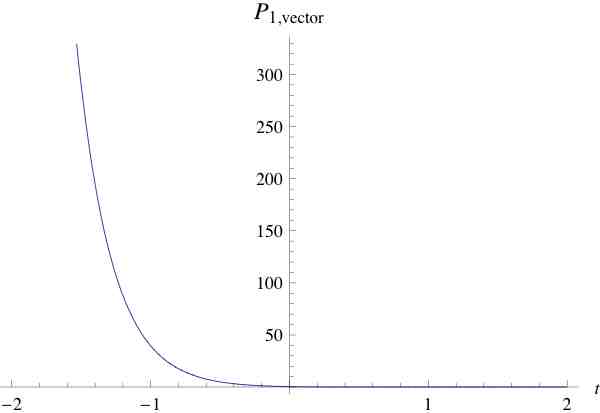}}} \\
\subfigure[$P_3^{1}$]{\label{emvectorsol2-c}\scalebox{0.70}{\includegraphics[width=7.5cm]{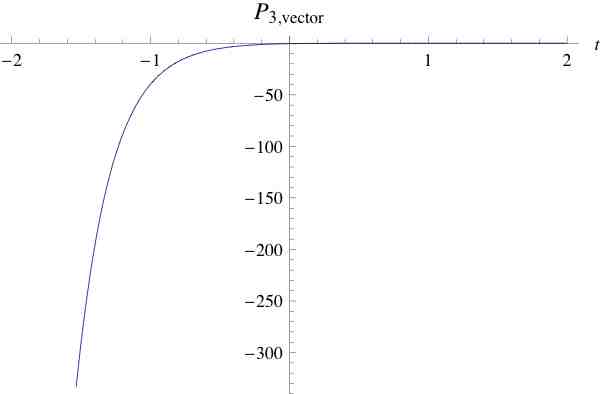}}} 
\end{center}
\caption{Energy density and pressure densities associated to the vector fields for the second solution.} 
\label{emvectorsol2}
\end{figure}

Constructing the corresponding four-dimensional solution, we find that the four-dimensional metric is of the form (\ref{formmetricscalefactors}), with the scale factors given by:
\begin{eqnarray} \label{scalefacsol2}
r^{[0]}(t)^2 & = & \frac{\rme^{\left(-3+2 \sqrt{30}\right) t} \sqrt{3+\rme^{4 t}}}{C[10]} \,, \nonumber \\
r^{[1]}(t)^2 & = &\frac{\rme^{\left(-3+\sqrt{30}\right) t} \sqrt{3+\rme^{4 t}}}{C[10]} \,, \nonumber \\
r^{[2]}(t)^2 & = & \frac{\rme^{3 t} C[10]}{\sqrt{3+\rme^{4 t}}}\,. 
\end{eqnarray}
These scale factors are plotted in figure \ref{plotsscalefactorssol2}.
In this case, the four-dimensional solution is characterized by non-trivial electric and magnetic fields. The non-zero components of the electric fields point in the $z$-direction and are given by:
\begin{eqnarray} \label{elfieldssol2}
\hat{\mathcal{F}}^0_{tz} & = & \frac{2 \sqrt{6} \rme^{\frac{1}{2} (8 t+C[3]-C[4]-C[5])} C[1] C[2] \sqrt{C[10]}}{\left(3+\rme^{4 t}\right)^2} \,, \nonumber \\
\hat{\mathcal{F}}^1_{tz} & = & \frac{2 \sqrt{6} \rme^{\frac{1}{2} (8 t+C[3]-C[4]-C[5])} C[1] \sqrt{C[10]}}{\left(3+\rme^{4 t}\right)^2} \,, \nonumber \\
\hat{\mathcal{F}}^2_{tz} & = & \frac{2 \sqrt{6} \rme^{\frac{1}{2} (8 t+C[3]-C[4]-C[5])} C[2] \sqrt{C[10]}}{\left(3+\rme^{4 t}\right)^2} \,, \nonumber \\
\hat{\mathcal{F}}^3_{tz} & = & -\frac{2 \sqrt{6} \rme^{\frac{1}{2} (8 t+C[3]-C[4]-C[5])} \sqrt{C[10]}}{\left(3+\rme^{4 t}\right)^2} \,,
\end{eqnarray}
while the magnetic fields point in the $z$-direction with constant strength:
\begin{eqnarray} \label{magfieldssol2}
\hat{\mathcal{F}}^0_{xy} & = & -\frac {2\sqrt {6} \rme^{\frac {1} {2} (-C[3] + C[4] + C[5])} C[16]} {\sqrt {C[10]}} \,, \nonumber \\
\hat{\mathcal{F}}^1_{xy} & = & 0 \,, \nonumber \\
\hat{\mathcal{F}}^2_{xy} & = & 0 \,, \nonumber \\
\hat{\mathcal{F}}^3_{xy} & = & 0 \,.
\end{eqnarray}
Plots of the electric fields can be found in figure \ref{plotselfieldssol2}.
Considering the total four-dimensional energy-momentum tensor, one finds the following expressions for the energy density and pressure densities:
\begin{eqnarray} \label{enpresstotsol2}
\rho^{tot}  & = & \frac{\rme^{\left(3-2 \sqrt{30}\right) t} \left(189+162 \rme^{4 t}+29 \rme^{8 t}\right) C[10]}{4 \left(3+\rme^{4 t}\right)^{5/2}} \,, \nonumber \\
P_1^{tot} & = & P_2^{tot}\  = \ \frac{\rme^{\left(3-2 \sqrt{30}\right) t} \left(189+162 \rme^{4 t}+29 \rme^{8 t}\right) C[10]}{4 \left(3+\rme^{4 t}\right)^{5/2}} \,, \nonumber \\
P_3^{tot} & = & \frac{\rme^{\left(3-2 \sqrt{30}\right) t} \left(189+66 \rme^{4 t}+29 \rme^{8 t}\right) C[10]}{4 \left(3+\rme^{4 t}\right)^{5/2}}\,.
\end{eqnarray}
The energy density and pressure densities that are associated to the scalar part of the energy-momentum tensor are given by:
\begin{eqnarray} \label{enpressscalarsol2}
\rho^{0}  & = &\frac{\rme^{\left(3-2 \sqrt{30}\right) t} \left(189+114 \rme^{4 t}+29 \rme^{8 t}\right) C[10]}{4 \left(3+\rme^{4 t}\right)^{5/2}} \,, \nonumber \\
P_1^{0} & = & P_2^{0}\  = \ P_3^{0}\ = \ \frac{\rme^{\left(3-2 \sqrt{30}\right) t} \left(189+114 \rme^{4 t}+29 \rme^{8 t}\right) C[10]}{4 \left(3+\rme^{4 t}\right)^{5/2}} \,,
\end{eqnarray}
while for the vector part, one obtains:
\begin{eqnarray} \label{enpressvectorsol2}
\rho^{1}  & = & \frac{12 \rme^{\left(7-2 \sqrt{30}\right) t} C[10]}{\left(3+\rme^{4 t}\right)^{5/2}} \,, \nonumber \\
P_1^{1} & = & P_2^{1}\  = \ \frac{12 \rme^{\left(7-2 \sqrt{30}\right) t} C[10]}{\left(3+\rme^{4 t}\right)^{5/2}} \,, \nonumber \\
P_3^{1} & = & -\frac{12 \rme^{\left(7-2 \sqrt{30}\right) t} C[10]}{\left(3+\rme^{4 t}\right)^{5/2}}\,.
\end{eqnarray}
We have plotted these quantities in figures \ref{emtotsol2}, \ref{emscalarsol2} and \ref{emvectorsol2}.

\begin{figure}[!h]
\begin{center}
\includegraphics[width=7.5cm]{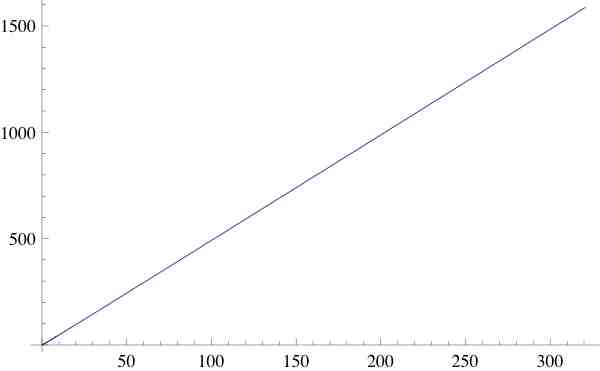}
\end{center}
\caption{Total volume for the solution of example 2.}
\label{totvolumeex2}
\end{figure}

We can again introduce the cosmic time $\tau$ as in (\ref{defcosmictime}). In this case, the cosmic time is given as a complicated hypergeometric function of $t$:
\begin{equation}
\tau = \frac{2 3^{1/4} e^{\left(-\frac{3}{2}+\sqrt{30}\right) t} \text{Hypergeometric2F1}\left[-\frac{1}{4},\frac{1}{8} \left(-3+2 \sqrt{30}\right),\frac{1}{8} \left(5+2 \sqrt{30}\right),-\frac{e^{4 t}}{3}\right]}{\left(-3+2 \sqrt{30}\right) \sqrt{C[10]}} \,.
\end{equation}
Considering the time evolution of the total volume $V(\tau)$ as in example 1, we again find that the total volume shows a linear behavior in $\tau$, as is shown in figure \ref{totvolumeex2}.

The previous two examples were such that the four-dimensional metric was diagonal. As can be seen from (\ref{redmetric}) and (\ref{solvectors3d}), this is due to the fact that the constant $\alpha$ is equal to zero in these examples. This situation is however not generic. The following example illustrates that solutions with non-diagonal four-dimensional metrics can also be addressed rather easily, using our methods.

\clearpage 

\subsubsection{Example 3} \label{sssec:ex3}

\begin{figure} [htp]
\begin{center}
\subfigure[$\log(\Delta(t))$]{\label{plotCSAsol3-a}\scalebox{0.80}{\includegraphics[width=7.5cm]{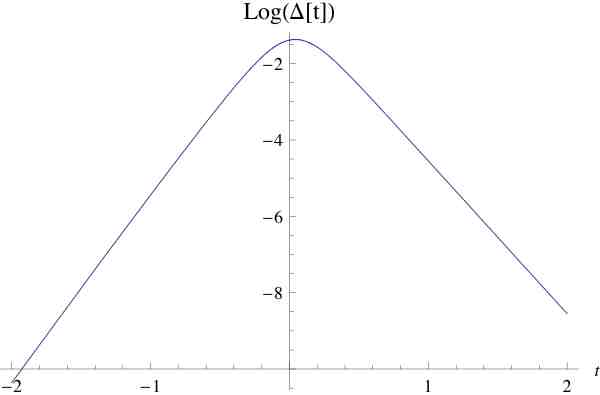}}} 
\subfigure[$\mathrm{h}^1(t)$]{\label{plotCSAsol3-b}\scalebox{0.80}{\includegraphics[width=7.5cm]{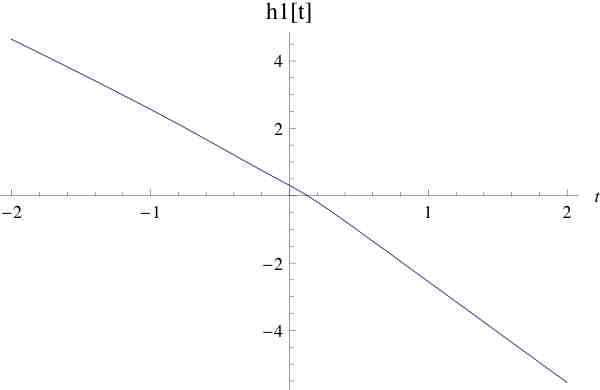}}} \\
\subfigure[$\mathrm{h}^2(t)$]{\label{plotCSAsol3-c}\scalebox{0.80}{\includegraphics[width=7.5cm]{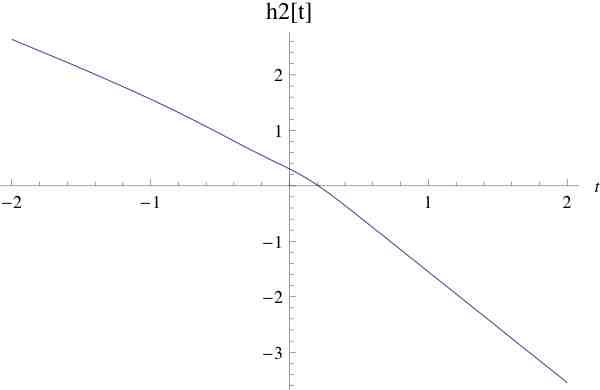}}} 
\subfigure[$\mathrm{h}^3(t)$]{\label{plotCSAsol3-d}\scalebox{0.80}{\includegraphics[width=7.5cm]{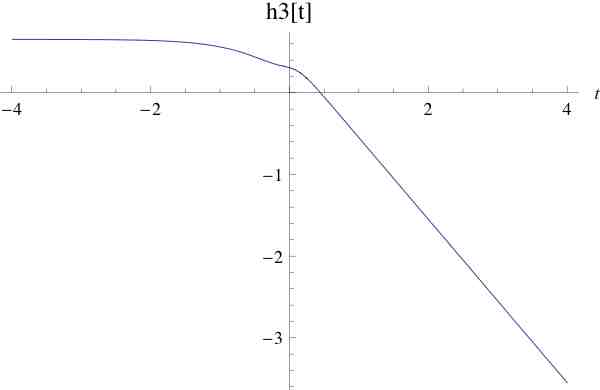}}} 
\end{center}
\caption{Cartan fields for the third solution.} 
\label{plotCSAsol3}
\end{figure}

For the third example, we have used the following values for the Euler angles (\ref{Eulerangles}):
\begin{equation} \label{anglesex3}
\theta_4 = \frac{\pi}{4} \,, \ \ \theta_8 = \frac{\pi}{3} \,, \qquad \theta_I = 0 \,, \quad I \neq 4, 8 \,.
\end{equation}
As the explicit solutions for the scalar fields are rather lengthy, we prefer to give only the solutions for the three-dimensional scalars associated to the Cartan generators:
{\renewcommand{\arraystretch}{1.4}
\begin{longtable}{|rcl|} 
\hline & & \\[-5mm]
$\Delta(t)$ & $=$ & $\displaystyle{\frac{\rme^{5 t} C[1]}{\sqrt{2+3 \rme^{2 t}+2 \rme^{8 t}+6\rme^{10 t}+3 \rme^{18 t}}}} \,, $ \\ [5mm]
$\mathrm{h}^1(t)$  & $=$ & $-2 t+C[2]+\frac{1}{2} \text{Log}\left[1+\rme^{8 t}\right]-\frac{1}{2} \text{Log}\left[2+3 \rme^{2 t}+3 \rme^{10 t}\right] \,,$ \\ [2mm]
$\mathrm{h}^2(t)$  & $=$ & $C[3]-2 \left(\frac{t}{2}-\frac{1}{4} \text{Log}\left[1+\rme^{8 t}\right]+\frac{1}{4} \text{Log}\left[2+3 \rme^{2 t}+3 \rme^{10 t}\right]\right)\,,$ \\ [2mm]
$\mathrm{h}^3(t)$  & $=$ & $C[4]+\frac{1}{2} \text{Log}\left[1+\rme^{8 t}\right]-\frac{1}{2} \text{Log}\left[2+3 \rme^{2 t}+3 \rme^{10 t}\right]\,.$ \\[2mm]
\hline
\caption{The solutions for the three-dimensional scalars associated to the Cartan generators with $c^0=1$, $c^1=2$, $c^3=3$, $c^4=4$ and $\theta_4=\pi/4$, $\theta_8=\pi/3$.} \label{sol3}
\end{longtable}} \vspace{-5mm}
Again, the asymptotic behavior of the Cartan fields, summarized in the following table, is noteworthy:
\begin{equation} \label{asympcsasol3}
\renewcommand{\arraystretch}{1.2}
\begin{array}{|c|c|}
\hline t \rightarrow -\infty & t \rightarrow + \infty \\ \hline 
\log(\Delta(t)) \sim 5 t & \log(\Delta(t)) \sim -4 t \\ 
\mathrm{h}^1(t)  \sim  -2 t & \mathrm{h}^1(t)  \sim  -3 t \\
\mathrm{h}^2(t)  \sim  -t & \mathrm{h}^2(t)  \sim -2 t \\ 
\mathrm{h}^3(t)  \sim  C[4] & \mathrm{h}^3(t)  \sim  - t \\ \hline
\end{array}
\end{equation}
Note that $\mathrm{h}^3(t)$ tends to a constant value for $t\rightarrow -\infty$.
To illustrate this asymptotic behavior, we have plotted the Cartan fields in figure \ref{plotCSAsol3}. The tendency of $\mathrm{h}^3(t)$ to become constant for $t\rightarrow -\infty$ is clearly visible.
\clearpage
For this specific example, the Weyl group elements $\sigma_{\pm \infty}$ that describe the above asymptotics, are both non-trivial. The element $\sigma_{+\infty}$ acts on the constants $c^i$ as:
\begin{equation} \label{actsigmamininfex3}
\setlength{\arraycolsep}{5mm}
\begin{array}{ll}\displaystyle{\sigma_{+\infty}(c^0)\  =  \ \frac{c^0}{2} - \frac{c^1}{2} - \frac{c^2}{2} - \frac{c^3}{2}} \,, &
\displaystyle{\sigma_{+\infty}(c^1)\  = \ -\frac{c^0}{2} + \frac{c^1}{2} - \frac{c^2}{2} - \frac{c^3}{2}} \,, \\[2mm]
\displaystyle{\sigma_{+\infty}(c^2)\  = \  -\frac{c^0}{2} - \frac{c^1}{2} + \frac{c^2}{2} - \frac{c^3}{2}} \,, & 
\displaystyle{\sigma_{+\infty}(c^3)\  = \ -\frac{c^0}{2} - \frac{c^1}{2} - \frac{c^2}{2} + \frac{c^3}{2}} \,.
\end{array}
\end{equation}
whereas the action of  $\sigma_{-\infty}$ is given by:
\begin{equation} \label{actsigmaplusinfex3}
\setlength{\arraycolsep}{5mm}
\begin{array}{ll}\displaystyle{\sigma_{-\infty}(c^0)\  =  \ \frac{c^0}{2} + \frac{c^1}{2} + \frac{c^2}{2} + \frac{c^3}{2}} \,, &
\displaystyle{\sigma_{-\infty}(c^1)\  = \ +\frac{c^0}{2} + \frac{c^1}{2} - \frac{c^2}{2} - \frac{c^3}{2}} \,, \\[2mm]
\displaystyle{\sigma_{-\infty}(c^2)\  = \  \frac{c^0}{2} - \frac{c^1}{2} + \frac{c^2}{2} - \frac{c^3}{2}} \,, & 
\displaystyle{\sigma_{-\infty}(c^3)\  = \ \frac{c^0}{2} - \frac{c^1}{2} - \frac{c^2}{2} + \frac{c^3}{2}} \,.
\end{array}
\end{equation}
One can again use these Weyl group elements and the explicit values (\ref{ciex}) to check the asymptotics of (\ref{asympcsasol3}).

\begin{figure} [!h]
\begin{center}
\subfigure[$\hat{\mathcal{F}}^1_{tx}$]{\label{plotelfieldsol3-a}\scalebox{0.70}{\includegraphics[width=7.5cm]{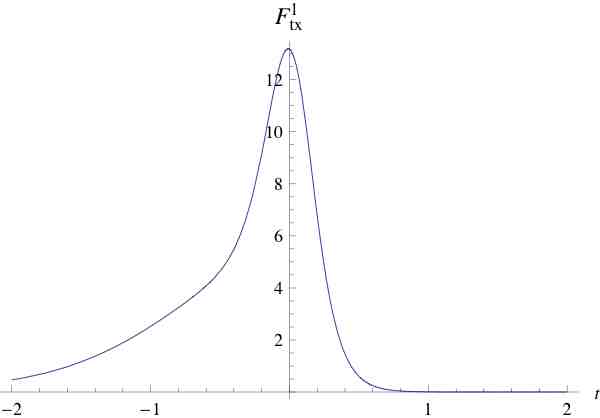}}} 
\subfigure[$\hat{\mathcal{F}}^1_{ty}$]{\label{plotelfieldsol3-b}\scalebox{0.70}{\includegraphics[width=7.5cm]{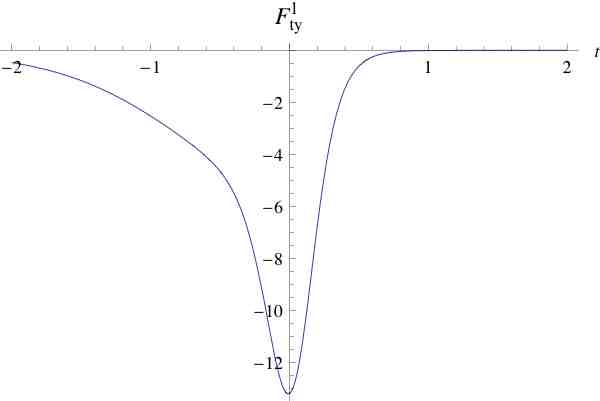}}} \\
\subfigure[$\hat{\mathcal{F}}^1_{tz}$]{\label{plotelfieldsol3-c}\scalebox{0.70}{\includegraphics[width=7.5cm]{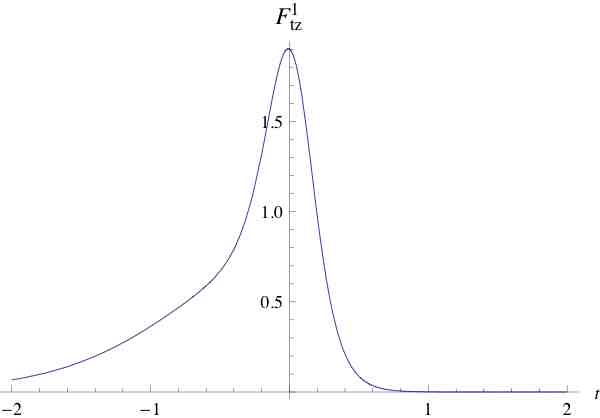}}} 
\subfigure[$\hat{\mathcal{F}}^2_{tx}$]{\label{plotelfieldsol3-d}\scalebox{0.70}{\includegraphics[width=7.5cm]{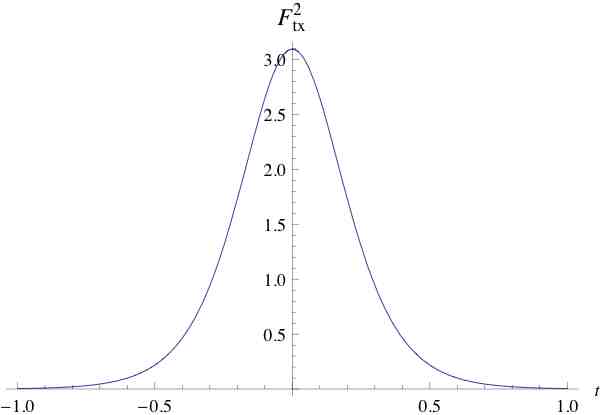}}} \\
\subfigure[$\hat{\mathcal{F}}^2_{ty}$]{\label{plotelfieldsol3-e} \scalebox{0.70}{\includegraphics[width=7.5cm]{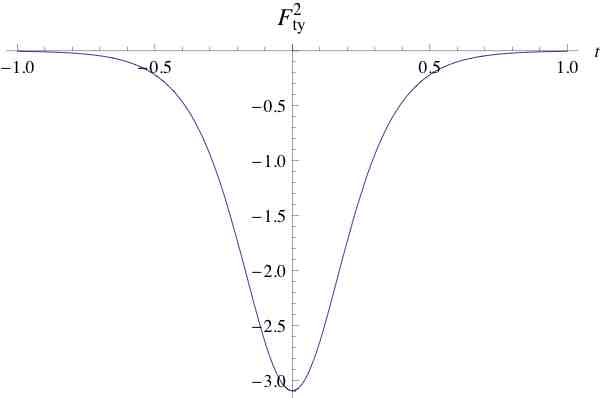}}} 
\subfigure[$\hat{\mathcal{F}}^2_{tz}$]{\label{plotelfieldsol3-f}\scalebox{0.70}{\includegraphics[width=7.5cm]{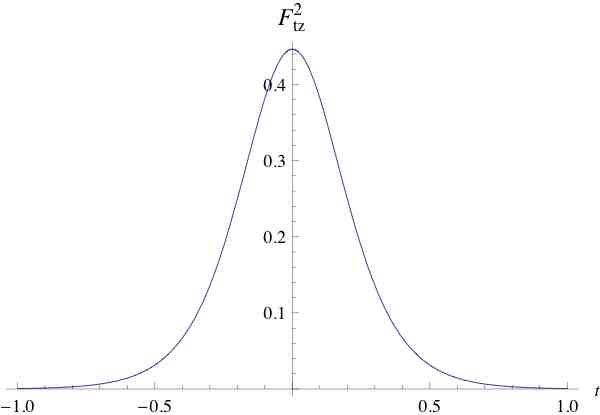}}} 
\end{center}
\caption{Electric fields for the third solution. The electric fields associated to the third and fourth vectors in the solution are proportional to the electric fields associated to the second vector.} 
\label{plotelfieldsol3}
\end{figure}

\begin{figure} [!h]
\begin{center}
\subfigure[$\hat{\mathcal{F}}^1_{xy}$]{\label{plotmagfieldsol3-a}\scalebox{0.70}{\includegraphics[width=7.5cm]{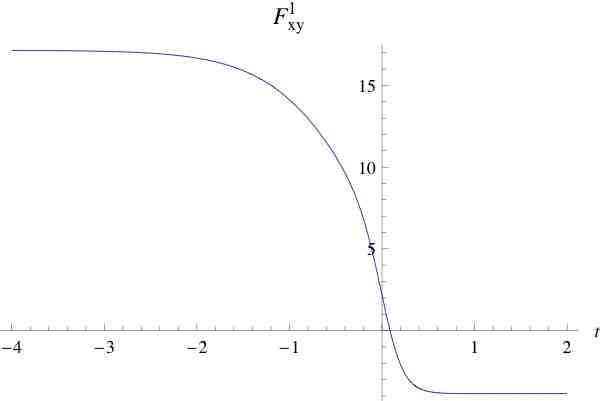}}} 
\subfigure[$\hat{\mathcal{F}}^2_{xy}$]{\label{plotmagfieldsol3-b}\scalebox{0.70}{\includegraphics[width=7.5cm]{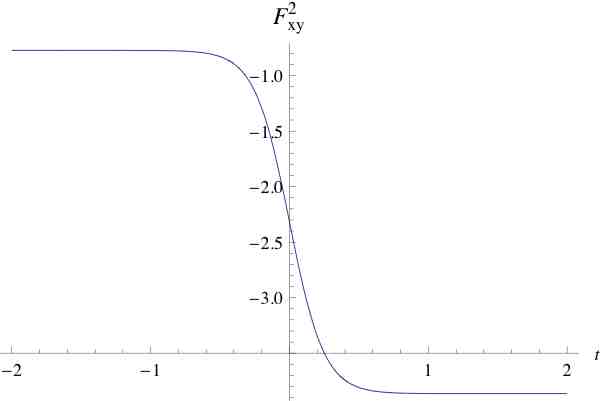}}} 
\end{center}
\caption{Magnetic fields for the third solution. The magnetic fields lie solely along the $z$-direction. The magnetic fields for the third and fourth vector in the solution are proportional to the magnetic field for the second vector.} 
\label{plotmagfieldsol3}
\end{figure}

The four-dimensional metric that corresponds to this solution is no longer diagonal. Its non-zero components are given by:
\begin{longtable}{rcl} \label{metricsol3}
$\hat{g}_{tt}$ & = & $\displaystyle{-\frac{\rme^{\left(-5+2 \sqrt{30}\right) t} \sqrt{\left(1+\rme^{8 t}\right) \left(2+3 \rme^{2 t} \left(1+\rme^{8 t}\right)\right)}}{C[1]}}$ \,, \nonumber \\[4mm]
$\hat{g}_{xx}$ & = & $\displaystyle{\frac{\rme^{-5 t} \left(\rme^{\sqrt{30} t} \left(1+\rme^{8 t}\right) \left(2+3 \rme^{2 t} \left(1+\rme^{8 t}\right)\right)+48 \rme^{10 t} y^2\right)}{\sqrt{\left(1+\rme^{8 t}\right) \left(2+3 \rme^{2 t} \left(1+\rme^{8 t}\right)\right)} C[1]}}$ \,. \nonumber \\[4mm]
$\hat{g}_{xy}$ & = & $\displaystyle{-\frac{48 \rme^{5 t} x y}{\sqrt{\left(1+\rme^{8 t}\right) \left(2+3 \rme^{2 t} \left(1+\rme^{8 t}\right)\right)} C[1]}}$ \,, \nonumber \\[4mm]
$\hat{g}_{xz}$ & = & $\displaystyle{\frac{4 \sqrt{3} \rme^{5 t} y}{\sqrt{\left(1+\rme^{8 t}\right) \left(2+3 \rme^{2 t} \left(1+\rme^{8 t}\right)\right)}}}$ \,, \nonumber \\[4mm]
$\hat{g}_{yy}$ & = & $\displaystyle{\frac{\rme^{-5 t} \left(\rme^{\sqrt{30} t} \left(1+\rme^{8 t}\right) \left(2+3 \rme^{2 t} \left(1+\rme^{8 t}\right)\right)+48 \rme^{10 t} x^2\right)}{\sqrt{\left(1+\rme^{8 t}\right) \left(2+3 \rme^{2 t} \left(1+\rme^{8 t}\right)\right)} C[1]}}$ \,, \nonumber \\[4mm]
$\hat{g}_{yz}$ & = & $\displaystyle{-\frac{4 \sqrt{3} \rme^{5 t} x}{\sqrt{\left(1+\rme^{8 t}\right) \left(2+3 \rme^{2 t} \left(1+\rme^{8 t}\right)\right)}}}$ \,, \nonumber \\[4mm]
$\hat{g}_{zz}$ & = & $\displaystyle{\frac{\rme^{5 t} C[1]}{\sqrt{2+3 \rme^{2 t}+2 \rme^{8 t}+6 \rme^{10 t}+3 \rme^{18 t}}}}$ \,.
\end{longtable}
Note however that for $t \rightarrow \pm \infty$, the off-diagonal components of the metric  (as well as $\hat{g}_{zz}$) tend to zero, due to the asymptotic behavior of $\Delta(t)$.  Considering the solutions for the electromagnetic field strengths, one now finds that there are also non-trivial electric fields in the $x$- and $y$-directions. The electric fields in the $z$-direction only depend on time, while the electric fields in the $x$- and $y$-direction also contain parts that linearly depend upon $y$ and $x$ respectively. All electric fields however tend to $0$ at $t = \pm \infty$. These electric fields are plotted in figure \ref{plotelfieldsol3}.
The magnetic fields are now no longer constant. There is a time-dependent magnetic field in the $z$-direction that interpolates between two different constant values at $\pm \infty$. Plots of the magnetic fields can be found in figure \ref{plotmagfieldsol3}.

\section{Ten-dimensional interpretation} \label{sec:10d}

As already mentioned, the $d=4$, $\mathcal{N}=2$ supergravity described by the special K\"ahler manifold (\ref{MSKisSL23}), can be obtained as a truncation of type IIB supergravity, compactified on a $K3 \times T^2/\mathbb{Z}_2$ orientifold (see e.g. \cite{Tripathy:2002qw,Andrianopoli:2003jf,Angelantonj:2003zx,D'Auria:2004qv,Dasgupta:2002ew}). In this section, we will re-interpret the four-dimensional solutions of section \ref{ssec:threeex} as solutions of ten-dimensional type IIB supergravity. 

\subsection{Three-dimensional supergravity solutions as type IIB solutions}

Let us consider the $K3 \times T^2/\mathbb{Z}_2$ orientifold compactification in some more detail. In the following, we will use indices $i$, $j = 1,2$ to denote the $T^2$-coordinates. Along $K3$, we will take complex coordinates $z^m$, $\bar{z}^{\bar{m}}$, $m=1,2$. As in the above discussion, we will continue to use $\hat{\mu}, \hat{\nu} = 0,\cdots,3$, as indices for the four non-compact coordinates. We will use capital latin letters $M, N$ to denote all ten-dimensional indices collectively.

The $\mathbb{Z}_2$-orientifold projection is generated by (see for instance \cite{Dasgupta:2002ew}):
\begin{equation} \label{deforientifoldproj}
\Omega \cdot (-1)^{F_L} \cdot \mathcal{I}_{45} \,,
\end{equation}
where $\Omega$ is the worldsheet orientation reversal, $(-1)^{F_L}$ is the operation that changes the sign of the left-moving space-time fermions and $\mathcal{I}_{45}$ is the orbifold projection that reverses the torus coordinates : $x^i \rightarrow - x^i$. The ten-dimensional fields transform in the following way under the action of $\Omega$ and $(-1)^{F_L}$:
\begin{equation} \label{orientrules}
\begin{array}{|c|c|c|c|}
\hline
\mathrm{field} & \Omega & (-1)^{F_L} & \Omega . (-1)^{F_L} \\ \hline
\mathrm{metric}\ g_{MN} & + & + & + \\
\mathrm{NS-}\mathrm{NS}\ 2\mathrm{-form}\ B^{(2)}_{MN} & - & + & - \\
\mathrm{RR}\ 2\mathrm{-form}\ C^{(2)}_{MN} & + & - & - \\
\mathrm{dilaton}\ \phi & + & + & + \\
\mathrm{RR\ axion}\ C^{(0)} & - & - & + \\
\mathrm{RR}\ 4\mathrm{-form}\ C^{(4)} & - & - & + \\ \hline
\end{array}
\end{equation}
Using this orientifold action, together with the Hodge diamond of $K3$:
\begin{equation} \label{hodgeK3}
\begin{array}{ccccc}
& & h^{0,0} & & \\
& h^{1,0} &  & h^{0,1} & \\
h^{2,0} &  & h^{1,1} &  & h^{0,2} \\
& h^{2,1} &  & h^{1,2} & \\
& & h^{2,2} & & \\
\end{array}\qquad  = \qquad \begin{array}{ccccc}
& & 1 & & \\
& 0 &  & 0 & \\
1 & & 20 &  & 1 \\
& 0 &  & 0 & \\
& & 1 & & \\
\end{array}\,,
\end{equation}
we can determine the four-dimensional low-energy spectrum of the $K3 \times T^2/\mathbb{Z}_2$ compactification.

The dilaton $\phi$ and the RR axion $C^{(0)}$ reduce trivially and lead to two real scalar fields in four dimensions. From (\ref{orientrules}), one sees that the orientifold truncation only keeps components of $B^{(2)}_{MN}$ and $C^{(2)}_{MN}$ that have one index along the $T^2$-directions. From the Hodge numbers (\ref{hodgeK3}), one can furthermore infer that the $K3\times T^2/\mathbb{Z}_2$-compactification of $B^{(2)}_{MN}$ and $C^{(2)}_{MN}$ only retains four four-dimensional vectors $B^{(2)}_{\hat{\mu} i}$ and $C^{(2)}_{\hat{\mu} i}$, $i=1,2$. After taking the self-duality condition into account, one similarly deduces that the reduction of the RR 4-form leads to 1 real scalar $C^{(4)}_{m\bar{n}p\bar{q}}$, that corresponds to the harmonic $(2,2)$-form on K3 and 22 real scalars $C^{(4)}_{ijmn}$, $C^{(4)}_{ij\bar{m}\bar{n}}$ and $C^{(4)}_{ijm\bar{n}}$, that correspond to the 22 harmonic two-forms on $K3$. Finally, the reduction of the metric leads to the metric $\hat{g}_{\hat{\mu}\hat{\nu}}$ in four dimensions. The metric excitations $g_{ij}$ along the torus lead to three scalar fields in four dimensions, one of which corresponds to the $T^2$-volume, while the other 2 parametrize the $T^2$ complex structure. The metric excitations along $K3$ correspond as usual to K\"ahler structure deformations and complex structure deformations. The K\"ahler structure deformations $\delta g_{m\bar{n}}$ are in one-to-one correspondence with harmonic $(1,1)$-forms, so they lead to 20 real degrees of freedom. The complex structure deformations on the other hand are of the form:
\begin{equation} \label{complstrdefK3}
\delta g_{mn} = \Omega_{mp} g^{p\bar{q}} \omega_{n \bar{q}} + (m \leftrightarrow n) \,,
\end{equation}
where $\Omega_{mn}$ corresponds to the holomorphic 2-form on $K3$ and $\omega_{m \bar{n}}$ is a closed $(1,1)$-form. Note however that the above formula gives zero when $\omega$ is given by the K\"ahler form of $K3$. We thus find that there are $h^{1,1} - 1 = 19$ independent complex structure deformations, that correspond to complex scalars in four dimensions. The moduli space of $K3$ metrics is thus 58-dimensional and can be shown to be given by the following symmetric space:
\begin{equation} \label{K3modspace}
\mathcal{M}_{K3} = \frac{\SO(3,19)}{\SO(3) \times \SO(19)} \times \mathbb{R}^+_{K3} \,,
\end{equation}
where the factor $\mathbb{R}^+_{K3}$ parametrizes the $K3$-volume. A summary of the bosonic four-dimensional spectrum of the $K3\times T^2/\mathbb{Z}_2$ reduction of type IIB supergravity can be found in table \ref{sumspectK3or}.
\begin{table}[!t]
\begin{center}
\renewcommand{\arraystretch}{1.4}
\begin{tabular}{|c|c|}
\hline metric & $\hat{g}_{\hat{\mu}\hat{\nu}}$ (1) \\ \hline
\multirow{2}{*}{scalars} & $\delta g_{mn}$ (38), $\delta g_{m\bar{n}}$ (20), $\delta g_{ij}$ (3), $\phi$ (1), $C^{(0)}$ (1), $C^{(4)}_{m\bar{n}p\bar{q}}$ (1), \\ 
 &  $C^{(4)}_{ijmn}$, $C^{(4)}_{ij\bar{m}\bar{n}}$, $C^{(4)}_{ijm\bar{n}}$ (22) \\ \hline
vectors & $B^{(2)}_{\hat{\mu} i}$ (2), $C^{(2)}_{\hat{\mu}i}$ (2) \\ \hline
\end{tabular}
\caption{Summary of the bosonic $d=4$ spectrum of type IIB on $K3\times T^2/\mathbb{Z}_2$. Between brackets, we have indicated the number of fields that appears after the reduction. For the scalars, the numbers between brackets refer to real degrees of freedom.}\label{sumspectK3or}
\end{center}
\end{table}
In total, this spectrum contains 86 scalar fields, as well as 4 vectors. The resulting theory corresponds to four-dimensional, $\mathcal{N}=2$ supergravity coupled to three vector multiplets and 20 hypermultiplets. The bosonic spectrum then indeed includes four vectors (one graviphoton and three vectors from the vector multiplets) and 86 scalar fields, of which 6 belong to the vector multiplets and 80 to the hypermultiplets. These scalars span a manifold that contains (\ref{K3modspace}) and that is the product of a special K\"ahler manifold $\mathcal{SK}$ and a quaternionic-K\"ahler manifold $\mathcal{QK}$:
\begin{eqnarray}
\mathcal{M} & = & \mathcal{SK} \times \mathcal{QK} \,, \nonumber \\
\mathcal{SK} & = & \frac{\SU(1,1)}{\U(1)} \times \frac{\SO(2,2)}{\SO(2) \times \SO(2)} \,, \nonumber \\
\mathcal{QK} & = & \frac{\SO(4,20)}{\SO(4) \times \SO(20)} \,.
\end{eqnarray}
The model considered in section \ref{ssec:threeex} thus corresponds to a $K3 \times T^2/\mathbb{Z}_2$ orientifold of type IIB supergravity where the hypermultiplets have been truncated. Of the 86 scalars in table \ref{sumspectK3or}, only the dilaton $\phi$, the RR axion $C^{(0)}$, one scalar $V_{K3}$ corresponding to the $K3$-volume, the scalar $C^{(4)}_{m\bar{n}p\bar{q}}$ and the $T^2$-complex structure belong to the vector multiplets. All other scalars belong to the hypermultiplets and are truncated. Thus in the following, we will assume that only the scalars of the vector multiplets are non-trivial. The six scalars of the vector multiplets are organized in three complex combinations, consistent with $d=4$, $\mathcal{N}=2$ supersymmetry. These complex scalars are often denoted as $S$, $T$ and $U$, hence the name $S$-$T$-$U$-model. Explictly, $S$, $T$ and $U$ are defined as:
\begin{eqnarray} \label{defstu}
S & = & S_1 + \rmi S_2 \ = \ C^{(0)} + \rmi \rme^{\phi} \,, \nonumber \\
T & = & T_1 + \rmi T_2 \ = \ \frac{g_{12}}{g_{22}} + \rmi \frac{\sqrt{g}}{g_{22}}\,, \qquad g =  \mathrm{torus \ metric} \,, \nonumber \\
U & = & U_1 + U_2 \ = \ U_1 + \rmi V_{K3} \,, \qquad \mathrm{where}\ C^{(4)} = U_1\ J \wedge J \,,
\end{eqnarray}
$J$ denoting the K\"ahler form on $K3$.

Let us now give the 10-dimensional fields in terms of the three-dimensional scalar fields. The ansatz for the ten-dimensional metric, dilaton $\phi$, RR axion $C^{(0)}$ and RR 4-form $C^{(4)}$ can be conveniently summarized by expressing the scalars $S$, $T$ and $U$ in terms of the three-dimensional fields. For the metric, this can be seen by the fact that the ansatz for the ten-dimensional metric depends on the four-dimensional scalars (\ref{defstu}) as follows:
\begin{equation} \label{ansatzmetric10d}
\rmd s_{10d}^2  =  V_{K3}^{-1/2} \rmd s_4^2 +  \frac{V_{K3}^{-1/2}}{T_2} \left[(\rmd v - T \rmd u)(\rmd v - \bar{T} \rmd u) \right]  + V_{K3}^{1/2} \rmd s^2_{K3} \,,
\end{equation}
where we have used real coordinates $u$, $v$ for the two-torus and $\rmd s^2_{K3}$ denotes the $K3$-metric. The scalars $S$, $T$ and $U$ depend in the following way on $\mathrm{h}^1(t)$, $\mathrm{h}^2(t)$, $\mathrm{h}^3(t)$, $\mathrm{g}^1(t)$, $\mathrm{g}^2(t)$ and $\mathrm{g}^3(t)$:
\begin{eqnarray} \label{stuin3dvars}
S & = & \mathrm{g}^3(t) + \rmi \rme^{-\mathrm{h}^3(t)} \,, \nonumber \\
T & = & \mathrm{g}^2(t) + \rmi \rme^{\mathrm{h}^2(t)}\,, \nonumber \\
U & = & -\frac{1}{\tilde{U}}\,, \qquad \mathrm{where}\ \tilde{U}  = \mathrm{g}^1(t) + \rmi \rme^{\mathrm{h}^1(t)} \,.
\end{eqnarray}
The two-forms on the other hand, can be expressed in terms of the four-dimensional vector fields:
\begin{eqnarray} \label{ansatztwoforms}
C^{(2)} & = & \sqrt{2} \hat{A}^0_{\hat{\mu}} \rmd x^{\hat{\mu}} \wedge \rmd u - \sqrt{2} \hat{A}^1_{\hat{\mu}} \rmd x^{\hat{\mu}} \wedge \rmd v \,, \nonumber \\
B^{(2)} & = & \sqrt{2} \hat{A}^2_{\hat{\mu}} \rmd x^{\hat{\mu}} \wedge \rmd u + \sqrt{2} \hat{A}^3_{\hat{\mu}} \rmd x^{\hat{\mu}} \wedge \rmd v \,.
\end{eqnarray}
The final expressions for the ten-dimensional solutions can be simplified by introducing the following one-form on the four-dimensional non-compact space-time:
\begin{equation} \label{defomega}
\Omega = -\frac{1}{2} y \rmd x + \frac{1}{2} x \rmd y \,.
\end{equation}
Note that $\Omega$ is not closed, rather it obeys:
\begin{equation} \label{domega}
\rmd \Omega = \rmd x \wedge \rmd y \,.
\end{equation}
Using (\ref{defstu}), (\ref{ansatzmetric10d}), (\ref{stuin3dvars}) and (\ref{ansatztwoforms}), we can obtain the formulas that express the ten-dimensional fields in terms of the three-dimensional scalar fields, summarized in table \ref{10solsin3dvars} \footnote{Note that we do not give the full solution for the RR four-form $C^{(4)}$. The solution given here should be supplemented with components along the four-dimensional space-time and $T^2$. These can however be determined by the components along $K3$ by requiring the field strength of $C^{(4)}$ to be self-dual.}.
\begin{longtable}{|rcl|} 
\hline & & \\[-1mm] $\displaystyle{\rmd s^2_{10d}}$ & = & $\displaystyle{\frac{\sqrt{\mathrm{g}^1(t)^2 + \rme^{2\mathrm{h}^1(t)}}}{\rme^{\frac{\mathrm{h}^1(t)}{2}}} \left[ -\frac{B^4}{\Delta} \rmd t^2 + \frac{B^2(t)}{\Delta(t)} (\rmd x^2 + \rmd y^2) + \Delta(t) (\rmd z + \alpha \Omega)^2 \right]}$  \\[5mm]
& & $\displaystyle{\frac{\sqrt{\mathrm{g}^1(t)^2 + \rme^{2\mathrm{h}^1(t)}}}{\rme^{\frac{\mathrm{h}^1(t)}{2}}} \left[\rme^{\mathrm{h}^2(t)} \left\{1 + \left(\frac{\mathrm{g}^2(t)}{\rme^{\mathrm{h}^2(t)}} \right)^2 \right\} \rmd u^2 + \rme^{-\mathrm{h}^2(t)} \rmd v^2 - 2 \frac{\mathrm{g}^2(t)}{\rme^{\mathrm{h}^2(t)}} \rmd u \rmd v \right]}$ \\[5mm] & & $\displaystyle{\frac{\rme^{\frac{\mathrm{h}^1(t)}{2}}}{\sqrt{\mathrm{g}^1(t)^2 + \rme^{2\mathrm{h}^1(t)}}} \rmd s^2_{K3}}$\,, \\[5mm]
$\displaystyle{\phi}$ & = & $\displaystyle{-\mathrm{h}^3(t)}$ \,, \\[5mm]
$\displaystyle{C^{(0)}}$ & = & $\displaystyle{\mathrm{g}^3(t)}$ \,, \\[5mm]
$\displaystyle{C^{(2)}}$ & = & $\displaystyle{\sqrt{2} \left(\beta^0 + \alpha \mathrm{p}^0(t)\right) \Omega \wedge \rmd u + \sqrt{2} \mathrm{p}^0(t) \rmd z \wedge \rmd u}$  \\[5mm]
&  & $-\displaystyle{\sqrt{2} \left(\beta^1 + \alpha \mathrm{q}^3(t)\right) \Omega \wedge \rmd v - \sqrt{2} \mathrm{q}^3(t) \rmd z \wedge \rmd v}$ \,, \\[5mm]
$\displaystyle{B^{(2)}}$ & = & $\displaystyle{\sqrt{2} \left(\beta^2 + \alpha \mathrm{q}^2(t)\right) \Omega \wedge \rmd u + \sqrt{2} \mathrm{q}^2(t) \rmd z \wedge \rmd u}$  \\[5mm]
&  & $+\displaystyle{\sqrt{2} \left(\beta^3 + \alpha \mathrm{p}^1(t)\right) \Omega \wedge \rmd v + \sqrt{2} \mathrm{p}^1(t) \rmd z \wedge \rmd v}$ \,, \\[5mm]
$\displaystyle{C^{(4)}}$ & = & $\displaystyle{-\frac{\mathrm{g}^1(t)}{(\mathrm{g}^1(t))^2 + \rme^{2 \mathrm{h}^1(t)}}} J \wedge J$ \,.\\[5mm] \hline
\caption{Summary of the ten-dimensional fields in terms of the three-dimensional scalar fields for the $K3 \times T^2/\mathbb{Z}_2$ orientifold compactification of type IIB supergravity. \label{10solsin3dvars}}
\end{longtable}
Note that the asymptotic behavior of these ten-dimensional cosmologies can be described in terms of the Weyl group of $\SO(4,4)$, in precisely the same way as described in section \ref{ssec:asympsol} for the four-dimensional solutions. More specifically, the Weyl group will determine the asymptotic time-dependent behavior of the metric, the dilaton and the RR four-form, as these depend on the Cartan scalars $\Delta(t)$, $\mathrm{h}^1(t)$, $\mathrm{h}^2(t)$, $\mathrm{h}^3(t)$. For the metric, the Weyl group thus explains how its various scale factors and the complex structure of the two-torus change in going from $-\infty$ to $+\infty$.

\subsection{Examples}

We can now reinterpret the solutions given in section \ref{ssec:threeex} as solutions of ten-dimensional type IIB supergravity. We will do this for the examples given in sections \ref{sssec:ex2} and \ref{sssec:ex3}. For reasons of clarity, we will however put some of the integration constants that appear in the second integration step, equal to zero. This is enough to show the general structure, as the solutions with all integration constants non-trivial often contain a lot of repetition. From table \ref{10solsin3dvars}, we can for instance see that the structure of $B^{(2)}$ and $C^{(2)}$ is very similar. We will thus choose the integration constants of the second integration step such that only $B^{(2)}$ or $C^{(2)}$ is non-trivial.

\subsubsection{Example 1}

As our first example, we consider the second example considered in section \ref{sssec:ex2}. For simplicity we will assume that
\begin{equation}
C[1] = C[2] = C[12] = C[13] = C[14] = C[15] = C[16] = 0 \,.
\end{equation}
In this case, the full 10-dimensional metric is given by:
\begin{longtable}{rcl}
$\displaystyle{\rmd s_{10d}^2}$ & = & $\displaystyle{-\frac{\rme^{-t+2 \sqrt{30} t+\frac{C[3]}{2}} \left(3+\rme^{4 t}\right)^{1/4}}{C[10]} \rmd t^2\text{ + }\frac{\rme^{\left(-1+\sqrt{30}\right) t+\frac{C[3]}{2}} \left(3+\rme^{4 t}\right)^{1/4}}{C[10]}(\rmd x^2 + \rmd y^2)}$ \\ & & $\displaystyle{+ \frac{\rme^{5 t+\frac{C[3]}{2}} C[10]}{\left(3+\rme^{4 t}\right)^{3/4}}\rmd z^2 + \rme^{3 t+\frac{C[3]}{2}+C[4]} \left(3+\rme^{4 t}\right)^{1/4}\rmd u^2 + \frac{\rme^{t+\frac{C[3]}{2}-C[4]}}{\left(3+\rme^{4 t}\right)^{3/4}}\rmd v^2}$  \\ & & $\displaystyle{+ \rme^{-2 t-\frac{C[3]}{2}} \left(3+\rme^{4 t}\right)^{1/4}\rmd s_{K3}^2}$ \,.
\end{longtable}
The RR two-form $C^{RR}_{(2)}$ is equal to zero, while the Kalb-Ramond field is given by
\begin{equation}
B^{NS}_{(2)} = \frac{\sqrt{3} \rme^{\frac{1}{2} (C[3]-C[4]-C[5])} \sqrt{C[10]}}{3+\rme^{4 t}} \rmd z \wedge \rmd v \,.
\end{equation} 
The dilaton $\phi$ and the RR axion $C^{(0)}$ are finally given by
\begin{equation}
\phi = -2 t-C[5]-\frac{1}{2} \text{Log}\left[3+\rme^{4 t}\right]\,, \qquad C^{(0)} = 0\,,
\end{equation}
while 
\begin{equation}
C^{(4)} = -\frac{\left(3+\rme^{4 t}\right) C[16]}{\rme^{8 t+2 C[3]}+\left(3+\rme^{4 t}\right) C[16]^2} J \wedge J \,.
\end{equation}

\subsubsection{Example 2}
Let us also consider the ten-dimensional interpretation of the example considered in section \ref{sssec:ex3}. Again, for the sake of simplicity, we will assume some integration constants to be zero:
\begin{equation}
C[5] = C[6] = C[7] = C[10] = C[12] = C[13] = 0 \,.
\end{equation}
The full ten-dimensional metric is then given by:
\begin{longtable}{rcl}
$\displaystyle{\rmd s_{10d}^2}$ & = &$\displaystyle{-\frac{\rme^{2 \left(-3+\sqrt{30}\right) t+\frac{C[2]}{2}} \left(1+\rme^{8 t}\right)^{1/4} \sqrt{\left(1+\rme^{8 t}\right) \left(2+3 \rme^{2 t} \left(1+\rme^{8 t}\right)\right)}}{\left(2+3 \rme^{2 t}+3 \rme^{10 t}\right)^{1/4} C[1]} \rmd t^2}$ \\ & & $\displaystyle{+ \frac{\rme^{-6 t+\frac{C[2]}{2}} \left(1+\rme^{8 t}\right)^{1/4} \left(\rme^{\sqrt{30} t} \left(1+\rme^{8 t}\right) \left(2+3 \rme^{2 t} \left(1+\rme^{8 t}\right)\right)+48 \rme^{10 t} y^2\right)}{\left(2+3 \rme^{2 t}+3 \rme^{10 t}\right)^{1/4} \sqrt{\left(1+\rme^{8 t}\right) \left(2+3 \rme^{2 t} \left(1+\rme^{8 t}\right)\right)} C[1]} \Big(\rmd x^2 + \rmd y^2\Big)}$ \\ & &$\displaystyle{+ \frac{\rme^{4 t+\frac{C[2]}{2}} \left(1+\rme^{8 t}\right)^{1/4} C[1]}{\left(2+3 \rme^{2 t}+3 \rme^{10 t}\right)^{1/4} \sqrt{2+3 \rme^{2 t}+2 \rme^{8 t}+6 \rme^{10 t}+3 \rme^{18 t}}}\rmd z^2}$ \\ & & $\displaystyle{-\frac{96 \rme^{4 t+\frac{C[2]}{2}} \left(1+\rme^{8 t}\right)^{1/4} x y}{\left(2+3 \rme^{2 t}+3 \rme^{10 t}\right)^{1/4} \sqrt{\left(1+\rme^{8 t}\right) \left(2+3 \rme^{2 t} \left(1+\rme^{8 t}\right)\right)} C[1]}\rmd x \rmd y}$ \\ & &  $\displaystyle{+ \frac{8 \sqrt{3} \rme^{4 t+\frac{C[2]}{2}} \left(1+\rme^{8 t}\right)^{1/4} y}{\left(2+3 \rme^{2 t}+3 \rme^{10 t}\right)^{1/4} \sqrt{\left(1+\rme^{8 t}\right) \left(2+3 \rme^{2 t} \left(1+\rme^{8 t}\right)\right)}}\rmd x \rmd z}$ \\ &&  $\displaystyle{-\frac{8 \sqrt{3} \rme^{4 t+\frac{C[2]}{2}} \left(1+\rme^{8 t}\right)^{1/4} x}{\left(2+3 \rme^{2 t}+3 \rme^{10 t}\right)^{1/4} \sqrt{\left(1+\rme^{8 t}\right) \left(2+3 \rme^{2 t} \left(1+\rme^{8 t}\right)\right)}}\rmd y \rmd z}$ \\ & &  $\displaystyle{+ \frac{\rme^{-2 t+\frac{C[2]}{2}+C[3]} \left(1+\rme^{8 t}\right)^{3/4}}{\left(2+3 \rme^{2 t}+3 \rme^{10 t}\right)^{3/4}}\rmd u^2 + \frac{\rme^{\frac{C[2]}{2}-C[3]} \left(2+3 \rme^{2 t}+3 \rme^{10 t}\right)^{1/4}}{\left(1+\rme^{8 t}\right)^{1/4}}\rmd v^2}$ \\ & & $\displaystyle{+ \frac{\rme^{t-\frac{C[2]}{2}} \left(2+3 \rme^{2 t}+3 \rme^{10 t}\right)^{1/4}}{\left(1+\rme^{8 t}\right)^{1/4}}\rmd s^2_{K3}}$ \,.
\end{longtable}
In this case, the Kalb-Ramond field $B^{(2)}$ is zero, whereas the RR two-form is given by:
\begin{eqnarray}
C^{(2)} & = & -\frac{4 \sqrt{2} \rme^{\frac{1}{2} (C[2]+C[3]+C[4])} y}{\left(2+3 \rme^{2 t}+3 \rme^{10 t}\right) \sqrt{C[1]}} \rmd x \wedge \rmd u + \frac{4 \sqrt{2} \rme^{\frac{1}{2} (C[2]+C[3]+C[4])} x}{\left(2+3 \rme^{2 t}+3 \rme^{10 t}\right) \sqrt{C[1]}} \rmd y \wedge \rmd u \nonumber \\ & &  + \frac{\sqrt{\frac{3}{2}} \rme^{\frac{1}{2} (4 t+C[2]+C[3]+C[4])} \left(1+\rme^{8 t}\right) \sqrt{C[1]}}{2+3 \rme^{2 t}+3 \rme^{10 t}} \rmd z \wedge \rmd u \,.
\end{eqnarray}
Furthermore, the RR four-form is zero in this case (due to the fact that it is proportional to $C[5]$), while the dilaton and RR axion are respectively given by:
\begin{equation}
\phi = -C[4]-\frac{1}{2} \text{Log}\left[1+\rme^{8 t}\right]+\frac{1}{2} \text{Log}\left[2+3 \rme^{2 t}+3 \rme^{10 t}\right] \,, \qquad C^{(0)} = 0 \,.
\end{equation}

\section{Conclusions and outlook}

In this paper, we have considered the problem of finding cosmological solutions of higher-dimensional ($d \geq 4$) supergravity theories. The method we used to find a large class of cosmologies consists in performing a dimensional reduction to three dimensions, where the theory reduces to a non-linear sigma model coupled to gravity. In case the target space of the resulting non-linear sigma model is a symmetric space, the field equations of the three-dimensional theory are integrable and can be solved for metrics of Friedmann-Lema\^{i}tre-Robertson-Walker (FLRW) type, via techniques that were developed earlier. We have described how the three-dimensional solutions can be uplifted to higher dimensions in an algorithmic manner via the process of dimensional oxidation. 

In this way, we obtain an interesting class of higher-dimensional cosmologies. In this paper, we were mainly concerned with finding cosmological solutions of $\mathcal{N}=2$, $d=4$ supergravity theories with only vector multiplets present. We have applied our algorithm to the so-called $S$-$T$-$U$ model, in which three vector multiplets are present and that stems from a $K3\times T^2/\mathbb{Z}_2$ orientifold compactification of type IIB supergravity. Some examples of explicit solutions have been given and their properties have been discussed. More specifically, we have devoted attention to the asymptotic behavior of these solutions and we have exhibited how the asymptotic states at $t=-\infty$ and $t=+\infty$ are related to each other via the action of the Weyl group of the three-dimensional duality algebra. We also showed how these solutions of the $S$-$T$-$U$ model can be seen as solutions of type IIB supergravity. Although the solutions in three dimensions are of FLRW-form, in higher dimensions they correspond to cosmologies where the algebra of translational isometries is non-abelian of the Heisenberg type.

Let us comment on some restrictions that were assumed in this paper, and how they can be removed. First of all, let us note that, although we mainly oxidized to four dimensions, this is not really a restriction. Indeed, for the class of $\mathcal{N}=2$ supergravities considered here, the highest dimension to which the three-dimensional theory can be oxidized is $d=5$ or even $d=6$ (see \cite{deWit:1991nm}, \cite{Fre:2006eu}, \cite{Andrianopoli:2004xu}). Similar uplifting formulas as described in this paper can then be used to perform the uplift to these higher dimensions. Indeed, we have for instance shown how to do the uplift to ten dimensions for the $S$-$T$-$U$ model. Secondly, the explicit example considered in this paper corresponded to a three-dimensional sigma model, whose symmetric target space is maximally non-compact. The theoretical discussion given in section \ref{sec:cosmsol} was however rather general and also holds for the more general case in which the three-dimensional sigma model is a non-maximally non-compact symmetric space. Finding time-dependent solutions in three dimensions can then be done by using the technique of Tits-Satake projections, as has been explained in \cite{Fre':2005sr}, \cite{Fre':2007hd} and \cite{Fre:2006eu}. The uplift of these solutions can then be performed along the general lines explained in section \ref{sec:cosmsol}. Thirdly, let us note that also the hypermultiplets can be included. The scalars of the hypermultiplets span a quaternionic-K\"ahler manifold by themselves and reduce trivially to three dimensions, resulting in a three-dimensional target space that is the direct product of two quaternionic-K\"ahler manifolds. The geodesic equations (\ref{3dgeodeq}) then decouple for both factors of this direct product and can be solved separately for each factor. Three-dimensional solutions can then be searched for in the usual manner and their uplifts can be considered.

Let us also note that in order to obtain more realistic cosmologies, we should not only look at ungauged supergravities, as was done in this paper, but we should consider gauged supergravities instead. These can for instance be obtained by considering flux compactifications and they exhibit non-trivial potential terms for the scalars. One then expects a much richer behavior, including cosmologies exhibiting inflationary periods and universes with accelerated expansion. It is therefore of obvious interest to see how our methods can also be adapted to the case of gauged supergravities. Finally, we would also like to mention that our solutions are restricted in the sense that they correspond to FLRW-cosmologies in three dimensions. In order to relax this restriction and to allow for different scale factors for all spatial dimensions, we would have to consider dimensional reductions to two dimensions and even to one single dimension. In these dimensions, the relevant duality algebras are no longer finite dimensional, but correspond to infinite-dimensional Kac-Moody algebras, as is well-known (see e.g. \cite{Julia:1981wc}, \cite{Julia:1985tw}, \cite{Nicolai:1991kx}). It is therefore an interesting problem to extend our methods to these cases. We hope to report on this in due course.

\section*{Acknowledgments.}

We acknowledge important and clarifying discussions with Prof. A. Sorin  
and also stimulating and very useful remarks from Prof. M. Porrati.

\section*{Appendix}
\addcontentsline{toc}{section}{Appendix}
\setcounter{section}{0}
\renewcommand{\theequation}{\Alph{section}.\arabic{equation}}
\renewcommand{\thesection}{\Alph{section}}
\renewcommand{\thetheorem}{\Alph{section}.\arabic{theorem}}

\section{The period matrix} \label{app:periodmatrix}
In this appendix, we collect the components of the symmetric matrix $\mathcal{N}_{IJ}$ that determines the couplings between scalars and vector fields in the four-dimensional supergravity Lagrangian. The non-zero independent components of the imaginary part of this matrix are given by:
\begin{longtable}{rcl}
$(\mathrm{Im} \mathcal{N})_{1,1}$ & = & $\displaystyle{-\frac{\rme^{\mathrm{h}^1(t)-\mathrm{h}^2(t)-\mathrm{h}^3(t)}}{\rme^{2 \mathrm{h}^1(t)}+\mathrm{g}^1(t)^2}} \,,$  \\
$(\mathrm{Im} \mathcal{N})_{1,2}$ & = & $\displaystyle{\frac{\rme^{\mathrm{h}^1(t)-\mathrm{h}^2(t)-\mathrm{h}^3(t)} \mathrm{g}^3(t)}{\rme^{2 \mathrm{h}^1(t)}+\mathrm{g}^1(t)^2}} \,,$  \\
$(\mathrm{Im} \mathcal{N})_{1,3}$ & = & $\displaystyle{\frac{\rme^{\mathrm{h}^1(t)-\mathrm{h}^2(t)-\mathrm{h}^3(t)} \mathrm{g}^2(t)}{\rme^{2 \mathrm{h}^1(t)}+\mathrm{g}^1(t)^2}} \,, $ \\
$(\mathrm{Im} \mathcal{N})_{1,4}$ & = & $\displaystyle{\frac{\rme^{\mathrm{h}^1(t)-\mathrm{h}^2(t)-\mathrm{h}^3(t)} \mathrm{g}^2(t) \mathrm{g}^3(t)}{\rme^{2 \mathrm{h}^1(t)}+\mathrm{g}^1(t)^2}} \,,$ \\
$(\mathrm{Im} \mathcal{N})_{2,2}$ & = & $\displaystyle{- \frac{\rme^{\mathrm{h}^1(t)-\mathrm{h}^2(t)-\mathrm{h}^3(t)} \big(\rme^{2 \mathrm{h}^3(t)} + \mathrm{g}^3(t)^2 \big)}{\rme^{2 \mathrm{h}^1(t)}+\mathrm{g}^1(t)^2}} \,,$  \\
$(\mathrm{Im} \mathcal{N})_{2,3}$ & = & $\displaystyle{-\frac{\rme^{\mathrm{h}^1(t)-\mathrm{h}^2(t)-\mathrm{h}^3(t)} \mathrm{g}^2(t) \mathrm{g}^3(t)}{\rme^{2 \mathrm{h}^1(t)}+\mathrm{g}^1(t)^2}} \,,$  \\
$(\mathrm{Im} \mathcal{N})_{2,4}$ & = & $\displaystyle{- \frac{\rme^{\mathrm{h}^1(t)-\mathrm{h}^2(t)-\mathrm{h}^3(t)} \mathrm{g}^2(t) \big(\rme^{2 \mathrm{h}^3(t)} + \mathrm{g}^3(t)^2 \big)}{\rme^{2 \mathrm{h}^1(t)}+\mathrm{g}^1(t)^2}} \,,$ \\
$(\mathrm{Im} \mathcal{N})_{3,3}$ & = & $\displaystyle{- \frac{\rme^{\mathrm{h}^1(t)-\mathrm{h}^2(t)-\mathrm{h}^3(t)} \big(\rme^{2 \mathrm{h}^2(t)} + \mathrm{g}^2(t)^2 \big)}{\rme^{2 \mathrm{h}^1(t)}+\mathrm{g}^1(t)^2}} \,,$ \\
$(\mathrm{Im} \mathcal{N})_{3,4}$ & = & $\displaystyle{- \frac{\rme^{\mathrm{h}^1(t)-\mathrm{h}^2(t)-\mathrm{h}^3(t)}  \big(\rme^{2 \mathrm{h}^2(t)} + \mathrm{g}^2(t)^2 \big)\mathrm{g}^3(t)}{\rme^{2 \mathrm{h}^1(t)}+\mathrm{g}^1(t)^2}} \,,$ \\
$(\mathrm{Im} \mathcal{N})_{3,3}$ & = & $\displaystyle{- \frac{\rme^{\mathrm{h}^1(t)-\mathrm{h}^2(t)-\mathrm{h}^3(t)}  \big(\rme^{2 \mathrm{h}^2(t)} + \mathrm{g}^2(t)^2 \big)\big(\rme^{2 \mathrm{h}^3(t)} + \mathrm{g}^3(t)^2 \big)}{\rme^{2 \mathrm{h}^1(t)}+\mathrm{g}^1(t)^2}} \,.$
\end{longtable}
For the real part of the period matrix, one has the following non-zero independent components:
\begin{longtable}{rcl}
$(\mathrm{Re} \mathcal{N})_{1,4}$ & = & $\frac{\mathrm{g}^1(t)}{\rme^{2 \mathrm{h}^1(t)} + \mathrm{g}^1(t)^2}$ \,, \nonumber \\
$(\mathrm{Re} \mathcal{N})_{2,3}$ & = & $\frac{\mathrm{g}^1(t)}{\rme^{2 \mathrm{h}^1(t)} + \mathrm{g}^1(t)^2}$ \,.
\end{longtable}

\section{The solvable algebra of $\SO(4,4)$} \label{app:solvalgsoconv}

Throughout the paper, the following conventions were used for the solvable algebra of $\SO(4,4)$. We have used the following form of the $\SO(4,4)$-invariant metric:
\begin{equation} \label{defetat}
\eta_t = \left( \begin{array}{cc} 0 & \omega_4 \\ \omega_4 & 0 \end{array} \right) \,,
\end{equation}
where $\omega_4$ is the matrix with entries 1 on the minor diagonal and all other entries zero:
\begin{equation} \label{defomega4}
\omega_4 = \left( \begin{array}{cccc} 0 & 0 & 0 & 1 \\ 0 & 0 & 1 & 0 \\ 0 & 1 & 0 & 0 \\ 1 & 0 & 0 & 0 \end{array}\right) \,.
\end{equation}
We then take a representation for which the solvable part of the algebra of $\SO(4,4)$ is generated by generators of the following type:
\begin{equation} \label{gensolvform}
\Lambda_t \in \mathrm{Solv}\Big(\frac{\SO(4,4)}{\SO(4) \times \SO(4)}\Big) \qquad \Leftrightarrow \qquad \Lambda_t = \left( \begin{array}{c|c} A & B \\ \hline 0 & -\omega_4 A^T \omega_4 \end{array} \right) \,,
\end{equation}
where $A$ is an upper triangular matrix
\begin{equation} \label{Auppertriang}
A = \left(\begin{array}{cccc} a_{11} & a_{12} & a_{13} & a_{14} \\ 0 & a_{22} & a_{23} & a_{24} \\ 0 & 0 & a_{33} & a_{34} \\ 0 & 0 & 0 & a_{44} \end{array}\right) \,,
\end{equation}
and $B$ obeys the condition:
\begin{equation} \label{restrictionB}
B^T \omega_4 + \omega_4 B = 0 \qquad  \Leftrightarrow  \qquad B = \left(\begin{array}{cccc} b_{11} & b_{12} & b_{13} & 0 \\ b_{21} & b_{22} & 0 & -b_{13} \\ b_{31} & 0 & -b_{22} & -b_{12} \\ 0 & -b_{31} & -b_{21} & -b_{11} \end{array}\right) \,.
\end{equation}
The matrix $\Lambda_t$ then obeys
\begin{equation}
\Lambda^T \eta_t + \eta_t \Lambda = 0 \,,
\end{equation}
and is upper triangular. Note that $A$ contains 10 parameters, while the matrix $B$, obeying the retriction (\ref{restrictionB}) contains 6 parameters. In total one thus obtains $16 = \mathrm{dim} \Big(\frac{\SO(4,4)}{\SO(4) \times \SO(4)}\Big)$ parameters.

Denoting by $E^{i,j}$ the $8\times 8$-matrix with $1$ on the $(i,j$)-th place and zero elsewhere, we have chosen the following representation for the Cartan generators in the Alekseevsky formalism (\ref{weightsVgeneral}):
\begin{eqnarray} \label{defCartansAl}
h_0 & = & \frac{1}{2}\left(E^{1,1} + E^{2,2} - E^{7,7} - E^{8,8} \right) \,, \nonumber \\
h_1 & = & \frac{1}{2}\left(E^{1,1} - E^{2,2} + E^{7,7} - E^{8,8} \right) \,, \nonumber \\
h_2 & = & \frac{1}{2}\left(E^{3,3} + E^{4,4} - E^{5,5} - E^{6,6} \right) \,, \nonumber \\
h_3 & = & \frac{1}{2}\left(E^{3,3} - E^{4,4} + E^{5,5} - E^{6,6} \right) \,.
\end{eqnarray}
For the positive roots $g_i$, $p_i$ and $q_i$, $i=0,\cdots,3$, the following representation was used:
\begin{equation} \label{defgpq}
\setlength{\arraycolsep}{15pt}
\begin{array}{lll}
g_0 = E^{\alpha_{9}} = E^{1,7} - E^{2,8}\,, & p_0 = E^{\alpha_{8}} = E^{1,6} - E^{3,8}\,, & q_0 = E^{\alpha_{4}} = E^{2,3} - E^{6,7}\,, \\ g_1 = E^{\alpha_{1}} = E^{1,2} - E^{7,8}\,, & p_1 = E^{\alpha_{2}} = E^{1,3} - E^{6,8}\,, & q_1 = E^{\alpha_{11}} = E^{2,6} - E^{3,7}\,, \\ g_2 = E^{\alpha_{12}} = E^{3,5} - E^{4,6}\,, & p_2 = E^{\alpha_{10}} = E^{2,5} - E^{4,7}\,, & q_2 = E^{\alpha_{3}} = E^{1,4} - E^{5,8}\,, \\ g_3 = E^{\alpha_{6}} = E^{3,4} - E^{5,6}\,, & p_3 = E^{\alpha_{5}} = E^{2,4} - E^{5,7}\,, & q_3 = E^{\alpha_{7}} = E^{1,5} - E^{4,8}\,.
\end{array}
\end{equation}
In order to determine in which order these positive roots should be exponentiated in e.g. formula (\ref{Eulerangles}), we have also indicated to which positive roots $\alpha_I$, $I=1,\cdots,12$ the generators $g_i$, $p_i$ and $q_i$ correspond. 

\section{Explicit construction of the Weyl group of $\SO(4,4)$} \label{app:weylso44}

The action (\ref{actweylgroup}) of the Weyl group of $\SO(4,4)$ on the constants $c^i$ in (\ref{C0expl}) can be explicitly constructed in the following manner. First, one constructs the 12 matrices
\begin{equation} \label{gensgenweyl}
\exp\left[\frac{\pi}{2} \left(E^{\alpha_I} - E^{\alpha_I \ T}\right)\right] \,,
\end{equation} 
where $E^{\alpha_I}$ correspond to the positive root generators of $\SO(4,4)$. 
By taking all possible products of these 12 matrices, one can close a discrete group $\mathcal{W}(\SO(4,4))$, which was called the 'generalized Weyl group' in \cite{Fre':2007hd}. Performing this procedure explicitly, shows that $\mathcal{W}(\SO(4,4))$ is a discrete subgroup of $\SO(4) \times \SO(4)$ containing 1536 elements. Using the explicit representation for the positive root generators in (\ref{defgpq}), one can moreover obtain an explicit $8 \times 8$-matrix representation of the elements of this generalized Weyl group. The generalized Weyl group thus obtained obeys the property that it leaves the non-compact Cartan subalgebra invariant upon conjugation:
\begin{equation} \label{actgenweyl}
\forall w \in \mathcal{W}\,,\ \forall \mathcal{C} \in \mathrm{CSA}^{\mathrm{nc}} \quad : \quad w \cdot \mathcal{C} \cdot w^T \in \mathrm{CSA}^{\mathrm{nc}} \,.
\end{equation}

The generalized Weyl group $\mathcal{W}(\SO(4,4))$ contains a discrete subgroup $N(\SO(4,4))$, consisting of all elements that stabilize the non-compact Cartan generator $\mathcal{C}$ :
\begin{equation} \label{defnormsubgr}
N(\SO(4,4)) = \left\{\gamma \in \mathcal{W}(\SO(4,4)) \quad : \quad \gamma \cdot \mathcal{C} \cdot \gamma^T = \mathcal{C}  \right\}\,.
\end{equation}
In the case at hand, the normal subgroup $N(\SO(4,4))$ consists of 8 elements. One can then divide the elements of $\mathcal{W}(\SO(4,4))$ in equivalence classes with respect to the normal subgroup $N(\SO(4,4))$, i.e. two elements $w_1$, $w_2$ $\in \mathcal{W}(\SO(4,4))$ belong to the same equivalence class when there exists an element $n \in N(\SO(4,4))$ such that
\begin{equation} \label{defequivclass}
w_1 = w_2 \cdot n \,.
\end{equation}
It turns out that the factor group of equivalence classes is isomorphic to the Weyl group $W(\SO(4,4))$ of $\SO(4,4)$:
\begin{equation} \label{defweyl}
\frac{\mathcal{W}(\SO(4,4))}{N(\SO(4,4))} \equiv W(\SO(4,4)) \,.
\end{equation}
By explicitly grouping the 1536 elements of $\mathcal{W}(\SO(4,4))$ in equivalence classes with respect to $N(\SO(4,4))$ and by choosing a representative in each class, we can thus easily obtain an explicit matrix representation of the Weyl group of $\SO(4,4)$. Note that the order of this Weyl group is given by $1536/8 = 192$, as expected.

The action of the $W(\SO(4,4))$ on the constants $c^i$ in (\ref{C0expl}) is then easily obtained by noting that:
\begin{equation} \label{cistrace}
c^i = \mathrm{Tr}\left(\mathcal{C}_0\cdot h_i \right)\,, \quad i=0,\cdots, 3 \,,
\end{equation}
and that $W(\SO(4,4))$ acts on $\mathcal{C}_0$ by conjugation as in (\ref{actgenweyl}). For $\sigma$ $\in$ $W(\SO(4,4))$, one thus concludes that the action of $\sigma$ on $c^i$ can be obtained via the following formula:
\begin{equation} \label{actweylc}
\sigma(c^i) = \mathrm{Tr}\left(\sigma \cdot \mathcal{C}_0 \cdot \sigma^T \cdot h_i \right)\,, \quad i=0,\cdots, 3 \,.
\end{equation}
It can be checked explicitly that this action of $W(\SO(4,4))$ on the constants $c^i$ is realized via orthogonal transformations, i.e.:
\begin{equation} \label{orthpropactweylc}
\sum_{i=0}^3 (c^i)^2 = \sum_{i=0}^3 \left[\sigma(c^i)\right]^2 \,.
\end{equation}


\providecommand{\href}[2]{#2}\begingroup\raggedright\endgroup

\end{document}